\documentclass{revtex4}
\usepackage{epsf}
\usepackage{amsmath}
\usepackage{amssymb}
\usepackage{bm}
\def\text{\mathrm}

\begin{document}           % End of preamble and beginning of text.
\title{Formation of molecules and entangled atomic
pairs from atomic BEC due to Feshbach resonance}

\author{Vladimir A. Yurovsky}

\affiliation{School of Chemistry, Tel Aviv University}

\date{15 March 2005}
\numberwithin{equation}{section}

\begin{abstract}Bose-Einstein condensate (BEC) is considered under
conditions of Feshbach resonance in two-atom collisions due to a
 coupling
of atomic pair and resonant molecular states. The association of
 condensate
atoms can form a molecular BEC, and the molecules can dissociate to
 pairs
of entangled atoms in two-mode squeezed states. Both entanglement and
squeezing can be applied to quantum measurements and information
processing.

The processes in the atom-molecule quantum gas are analyzed using two
theoretical approaches. The mean-field one takes into account
 deactivating
collisions of resonant molecules with other atoms and molecules,
 neglecting
quantum fluctuations. This method allows analysis of inhomogeneous
 systems,
such as expanding BEC. The non-mean-field approach --- the parametric
approximation --- takes into account both deactivation and quantum
fluctuations. This method allows determination of optimal conditions
 for
formation of molecular BEC and describes Bose-enhanced dissociation of
molecular BEC, as well as entanglement and squeezing of the
 non-condensate
atoms.\end{abstract}

\maketitle

\section*{Introduction}

Recent developments in the field of Bose-Einstein condensation of
 dilute atomic
gases are described in review articles
 \cite{PW98,DGPS99,CBY01,RCNB04,Yukalov04} and a
book \cite{PS03}. One of the last accomplishments in this field is
 the formation of
molecular Bose-Einstein condensate (BEC) from an atomic one
\cite{HKMWCNG03,MKHCNG04,DVMR03,DVMR04,XMACMK03,MAXCK04} and from a
 quantum-degenerate
Fermi gases
\cite{JILARTBJ03,JILAGRJ03,RiceSPH03,RiceC03,ENSB04,InnsJ03,InnsJ03b,MITZ03,MITK04}.
These experiments used the effect of Feshbach resonance (see Ref.\
 \cite{TTHK99}),
appearing in multichannel scattering when the collision energy of an
 atomic pair lies
in the vicinity of the energy of a bound (molecular) state in a
 closed channel. An
analog of this effect --- the confinement induced resonance --
 appears in scattering
under tight cylindrical confinement (see Refs.\
 \cite{O98,BMO03,MBO04}). A cooperative
effect of the confinement and closed channels has been considered in
 Ref.\ \cite{Y05}.

Feshbach resonance allows control of  BEC properties by tuning the
elastic scattering length, as has been proposed in Refs.\
\cite{Tiesinga,Fedichev1996a,Bohn1997a}. The open and closed channels
 can
be coupled by resonant optical fields \cite{Fedichev1996a,Bohn1997a}
 or by
hyperfine interaction \cite{Tiesinga}, using the Zeeman effect for the
energy detuning control. A coherent formation of molecules due to such
coupling schemes has been considered in Refs.\
\cite{TTHK99,JBBS98,Javanainen1998,DKH98,TTCHK99}. The experimental
realization of the scattering length control
 \cite{IASMSK98,SIAMSK99,C00}
has demonstrated a drastic condensate loss when the resonance was
approached. This loss has been attributed to two loss mechanisms: a
deactivation of the temporarily formed molecules in inelastic
 collisions
with atoms and molecules (see Refs.\ \cite{TTHK99,YBJW99,AV99}) and a
dissociation of the molecules resulting in the formation of
 non-condensate
atoms (see Refs.\ \cite{AV99,MTJ00,YBJW00,GGR01,PM01,HPW01}).
 Secondary
collisions of the relatively hot deactivation products can lead to an
additional loss (see Refs.\ \cite{YBJW99,GS99,SMASRB01}).

Many-body effects on the formation of molecules can be described in
a mean-field approach by a set of coupled Gross-Pitaevskii equations
\cite{GP} for the atomic and molecular mean fields (see Ref.\
\cite{TTHK99}). The coupled fields form a hybrid atom-molecule
condensate. Deactivation losses can be introduced into this approach
 by
adding imaginary non-linear terms. A derivation of such equations is
presented in Sec.\ \ref{SecMeanField} below. The mean field approach
 has
been used for description of atom-molecule oscillations
\cite{TTHK99,TTCHK99,Heinzen2000a}, allowing analytical solutions for
some time-dependent models \cite{IMCGJ04,ICN04}. This approach
 describes
also the effects of self-trapping and soliton formation
\cite{Cusack2001a,VKD04}, vorticity \cite{AOKJ02}, and expansion
\cite{YB04} of hybrid atom-molecule condensates.

The second loss mechanism --- the dissociation of molecules
onto non-condensate atoms --- has been described by a two-body
theory in Ref.\ \cite{MTJ00} and introduced as additional loss terms
into the many-body mean-field equations in Ref.\ \cite{YBJW00}. Both
these approaches forbid the dissociation in a backward sweep, when
the molecular state crosses the atomic ones in a downward direction.
This manner, proposed in Ref.\ \cite{MTJ00}, has been used in
experiments \cite{HKMWCNG03}-\cite{MAXCK04}. However, the
dissociation can proceed even in a backward sweep, and correct
description of this process requires more comprehensive theories. On
a two-body level it can be described as a so-called
``counterintuitive'' transition \cite{counter_int}. Apart from this,
the dissociation of molecular BEC demonstrates quantum many-body
effects beyond the mean-field approach, such as Bose-enhancement
\cite{YB03,YB03j}. The effects of deactivation and dissociation
losses are non-additive (see Ref.\ \cite{YBJW00}).

The atomic pairs produced by dissociation are formed in two-mode
squeezed states which are entangled \cite{YB03}. Squeezed states are
characterized by noise reduction, and can be applied in
 communications and
measurements. Two-mode squeezed states have the property of relative
 number
squeezing (see Ref.\ \cite{RGB02}), which is important for atomic
interferometry. Entangled states of a decomposable system cannot be
 expressed
as a product of the component states, and can be used in quantum
 computing
and communications (see Refs.\ \cite{DBB02,DB04}.) An entanglement
 measure
has been considered in Refs.\ \cite{YMPL03,YPRL03}. Some aspects of
 the
entanglement have been considered within two-body theory of
 dissociation of
individual molecules in Ref.\ \cite{OK01}, but the analysis of
 many-body
effects requires a theory treating the non-condensate atoms as quantum
fluctuations.

Several theoretical methods are available for a more correct
treatment of the many-body non-mean-field effects. A numerical
solution of stochastic differential equations has been used in
this context in Refs.\
\cite{PM01,HOP01,Hope01,HO01,OPC01,KD02,KOD04,SSK06}. The role of
quantum statistics of the initial state has been investigated by
this method in Refs.\ \cite{OP03,OP04,Olsen04,OBC04}. A direct
solution of the many-body Schr\"odinger \cite{Javanainen1999a} and
Liouville von-Neumann equations \cite{VYA01} has been applied only
to a small number of atomic modes (maximum two, in Ref.\
\cite{MV02}). A special time-independent case of a single atomic
mode allows an exact solution in terms of the algebraic Bethe
ansatz \cite{LZMG03}.  Apart from this, non-mean field effects of
a periodic potential \cite{M04} and of the directional emission of
correlated atomic pairs \cite{VM02} have been analyzed.

The prominent many-body quantum effects, such as entanglement,
squeezing, and Bose-enhancement, can de described by methods taking
 into
account second-order correlations. In the Hartree-Fock-Bogoliubov
formalism of Refs.\ \cite{HPW01,KH02}, the mean field equations are
complemented by equations for the normal and anomalous densities of
 atomic
fluctuations. The macroscopic quantum dynamics approach \cite{KB02}
considers the molecules as two-atom bound states, incorporating actual
interatomic potentials. The system dynamics is described by an
 infinite
set of equations for non-commutative cummulants. The first-order
 cummulant
approach, applied to atom-molecule systems in Refs.\
\cite{KGB03,GKGB04,KGG04,GKGTJ04,GKB05}, keeps the cummulants
corresponding to mean fields and anomalous densities. This truncation
allows to express the contribution of quantum fluctuations in terms of
two-body transition matrix. The resulting dynamic equations are
 simpler
then the Hartree-Fock-Bogoliubov ones, allowing analyses of spatially
inhomogeneous systems. The truncation neglecting the cummulants
corresponding to normal density is justified under conditions of
 majority
of current experiments. However, it leads to the loss of some quantum
effects, such as Bose-enhancement (see Sec.\ \ref{SecNorDen}). These
effects could be taken into account in higher-order cummulant
 approaches.
Such approach incorporating third-order correlations has been
 developed in
Ref.\ \cite{Koehler02} for pure atomic gas.

The theoretical approaches of Refs.\
\cite{PM01,HPW01,HOP01}-\cite{GKB05} did not take into account the
 effects
of deactivating collisions. The present chapter describes the
 parametric
approximation --- a non-mean-field formalism in which the damping
 effects
are incorporated (see Sec.\ \ref{SecParAppr}). This approach has been
applied to the case of a single atomic mode in Refs.\
 \cite{VYA01,YBJ02},
where analytical solutions have been obtained (see also Sec.\
\ref{SecExSol} and Ref.\ \cite{KS03}), to the formation of
 instabilities
in pure atomic BEC in Ref.\ \cite{Y02}, and to multimode atom-molecule
quantum gases in Refs.\ \cite{YB04,YB03,YB03j}. When the deactivating
collisions are neglected, the parametric approximation becomes
 equivalent
to the Hartree-Fock-Bogoliubov approach of Refs.\ \cite{HPW01,KH02}
(see
Sec.\ \ref{SecHFB}, where Hartree-Fock-Bogoliubov equations involving
deactivation are also derived).

This chapter presents applications of the mean field and
parametric approximations to the description of condensate losses
(Sec.\ \ref{SecLoss}), formation of molecules (Sec.\
\ref{SecForMol}), and production of entangled atoms (Sec.\
\ref{SecEntAt}). The results demonstrate the importance of both
non-mean-field and dumping effects for a correct description of
processes in atom-molecule quantum Bose gases. A different
situation takes place in the case of molecular BEC formed from
Fermi atoms, where the deactivation can be neglected due to Pauli
blocking (see Refs.\ \cite{P03,PSS04,PSS05}). For a theory of
molecular formation from quantum-degenerate Fermi gases, which is
not presented here, see Refs.\ \cite{PVB04,CGKJ04,TV04}.
Production of entangled fermionic pairs has been analyzed in 
Ref.\ \cite{Kheruntsyan06}.

A system of units in which Planck's constant is $\hbar =1$ is used
 below.

\section{Second-quantized description of hybrid atom-molecule
gases\label{SecQuan}}

\begin{figure}
\epsfxsize=0.6\textwidth  \epsfbox{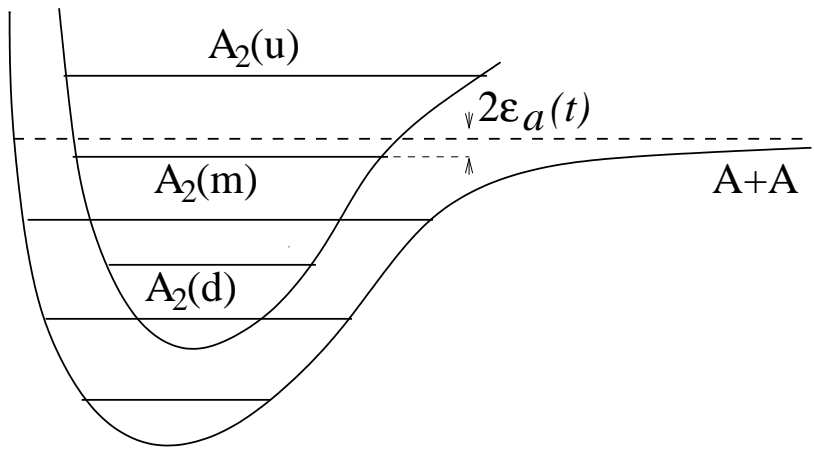}

\caption{Schematic description of channel potentials
and molecular states. The dashed line marks the open channel
threshold. \label{FigFeshbach}}

\end{figure}

The effect of Feshbach resonance appears when the
collision energy of a pair of similar atoms of type A in an
open channel is close to the energy of a bound state A$_{2}\left(
 m\right) $
in the closed channel (see Fig.\ \ref{FigFeshbach}). The
temporary formation and dissociation of the resonant
(Feshbach) molecular state A$_{2}\left( m\right) $ can be described
 by the
reversible reaction
\begin{equation}
\text{A + A  }\rightleftarrows\text{ A}_{2}\left( m\right)  .
 \label{RCol}
\end{equation}
Although a resonance can be related to several closely-spaced
bound states, this chapter considers a case of a non-degenerate
resonant state, well separated from other bound states. This state is
described by the field annihilation operator $\hat{\psi }_{m}\left(
 {\bf r}\right) $, while the atoms
are described by the operator $\hat{\psi }_{a}\left( {\bf r}\right)
 $. The Hamiltonian for the free
fields can be written in the form
\begin{equation}
\hat{H}_{\text{free}}=\int d^{3}r\hat{\psi }^{\dag }_{a}\left( {\bf
 r}\right) \hat{H}_{a}\hat{\psi }_{a}\left( {\bf r}\right) +\int
 d^{3}r_{m}\hat{\psi }^{\dag }_{m}\left( {\bf r}_{m}\right)
 \hat{H}_{m}\hat{\psi }_{m}\left( {\bf r}_{m}\right)  ,
\end{equation}
where
\begin{equation}
\hat{H}_{a}={1\over 2m}\hat{{\bf p}}^{2}+V_{a}\left( {\bf r}\right)
+\epsilon _{a},\qquad \hat{H}_{m}={1\over 4m}\hat{{\bf p}}^{2}_{m}
+V_{m}\left( {\bf r}_{m}\right)  \label{HaHm}
\end{equation}
\medskip
are the Hamiltonians for the noninteracting atoms and
molecules in the representation of first quantization. Here $m$
is the atom mass, $V_{a}\left( {\bf r}\right) $ and $V_{m}\left( {\bf
 r}_{m}\right) $ are, respectively, the
atomic and molecular trap potentials as functions of the
position ${\bf r}$ of the atomic (or ${\bf r}_{m}$  of the molecular)
 center of
mass. The time-dependent Zeeman shift
\begin{equation}
\epsilon _{a}\left( t\right) =-{1\over 2}\mu \left( B\left( t\right)
 -B_{0}\right)
\end{equation}
of the atom in an external magnetic field $B\left( t\right) $ is
 counted
from half the energy of the resonant molecular state (which is
fixed as the zero energy point). Here
\begin{equation}
\mu =2\mu _{a}-\mu _{m} ,
\end{equation}
$\mu _{a}$  and $\mu _{m}$  are the magnetic moments of the atom and
 the
resonant molecule, respectively, and $B_{0}$  is the resonance value
of $B$ in the free space (when $V_{a}=V_{m}=0$).

Since the atoms and molecules are treated here as
independent particles, the interaction responsible for the
atom-molecule coupling [reaction (\ref{RCol})] can be
written in the general form
\begin{equation}
\hat{V}_{h}=\int d^{3}r d^{3}r^\prime  V_{h}\left( {\bf r}-{\bf
 r}^\prime \right)  \hat{\psi }^{\dag }_{m}\left( {{\bf r}+{\bf
 r}^\prime \over 2}\right) \hat{\psi }_{a}\left( {\bf r}\right)
 \hat{\psi }_{a}\left( {\bf r}^\prime \right)  , \label{Vh}
\end{equation}
in which the molecule preserves the position of the
center of mass. Although the interaction is localized within a
range of atomic size, negligibly small compared to the
condensate size and the relevant de Broglie wavelengths, the
use of a zero-range interaction would lead to a divergence in
the ensuing calculations. Therefore we keep this interaction
as a finite-range function $V_{h}\left( {\bf r}-{\bf r}^\prime \right
) $ of the interatomic
distance.

The resonance bound state A$_{2}\left( m\right) $ is generally an
 excited
rovibrational state, belonging to non-ground states of the
fine and hyperfine structures (see Fig.\ \ref{FigFeshbach}).
This state can be deactivated by an exoergic collision with a
third atom of the condensate \cite{TTHK99,YBJW99,AV99,YBJW00},
\begin{equation}
\mathrm{A}_{2}(m)+\text{A}\rightarrow \text{A}_{2}(d)+\text{A  ,
 }\label{AMCol}
\end{equation}
bringing the molecule down to a lower state A$_{2}\left( d\right) $
(see
Fig.\ \ref{FigFeshbach}), while releasing kinetic energy to
the relative motion of the reaction products. Although the
collision occurs with a vanishingly small kinetic energy,
rates of such inelastic processes remain finite at near-zero
energies \cite{Forrey99,SCHHL02,MAXCK04}. A variant of this
process, involving deactivation by a collision with another
molecule (rather than an atom), of the type
\begin{equation}
\mathrm{A}_{2}(m)+\mathrm{A}_{2}(m)\rightarrow \text{A}_{2}\left(
 d\right) +\text{A}_{2}\left( u\right)  . \label{MMCol}
\end{equation}
would require a significant molecular density to be
effective. The two molecules emerge in two states A$_{2}\left(
 d\right) $ and
A$_{2}\left( u\right) $  (see Fig.\ \ref{FigFeshbach}), where
 A$_{2}\left( u\right) $ can be a
bound molecular state above A$_{2}\left( d\right) $, or a continuum
 state of a
dissociating molecule. The reaction can take place as long as
the corresponding internal energies $E_{d}$  and $E_{u}$   obey the
inequality $E_{d}+E_{u}\le 0$. A particularly effective reaction of
 type
(\ref{MMCol}) would occur in the near-resonant case, in which
$0<E_{u}<|E_{d}|$. A typical example, common in VV-relaxation, is that
of $v+v\rightarrow \left( v+1\right) +\left( v-1\right) $ (where $v$
 is the vibrational quantum number
of the state $m$). In this example, the kinetic energy is
provided by the vibrational anharmonicity.

The products of reactions (\ref{AMCol}) and (\ref{MMCol})
are described by the annihilation operators $\hat{\psi }_{d}\left(
 {\bf r}\right) $ and $\hat{\psi }_{u}\left( {\bf r}\right) $,
respectively, for each of the states A$_{2}\left( d\right) $ and
 A$_{2}\left( u\right) $ . The
Hamiltonians of free fields can be represented as
\begin{equation}
\hat{H}_{d,u}=\int d^{3}r_{m}\hat{\psi }^{\dag }_{d,u}\left( {\bf
 r}_{m}\right) \left( {1\over 4m}\hat{{\bf p}}^{2}_{m}+E_{d,u}\right)
 \hat{\psi }_{d,u}\left( {\bf r}_{m}\right)  ,
\end{equation}
where the trap potentials and Zeeman shifts can be
neglected compared to the transition energies $E_{d,u}$. The
deactivating collisions (\ref{AMCol}) and (\ref{MMCol}) are
described, respectively, by the following interactions
\begin{eqnarray}
\hat{V}_{d}=\int d^{3}rd^{3}r_{m}d_{d}\left( |{\bf r}-{\bf r}_{m}
|\right) \hat{\psi }^{\dag }_{a}\left( {\bf r}\right) \hat{\psi
 }^{\dag }_{d}\left( {\bf r}_{m}\right) \hat{\psi }_{a}\left( {{\bf
 r}+2{\bf r}{ } _{m}\over 3}\right) \hat{\psi }_{m}\left( {{\bf r}
+2{\bf r}{ } _{m}\over 3}\right)  \label{Vd}
\\
\hat{V}_{ud}=\int d^{3}r_{1}d^{3}r_{2}d_{ud}\left( |{\bf r}_{1}-{\bf
 r}_{2}|\right) \hat{\psi }^{\dag }_{u}\left( {\bf r}_{1}\right)
 \hat{\psi }^{\dag }_{d}\left( {\bf r}_{2}\right) \hat{\psi
 }_{m}\left( {{\bf r}_{1}+{\bf r}{ } _{2}\over 2}\right) \hat{\psi
 }_{m}\left( {{\bf r}_{1}+{\bf r}{ } _{2}\over 2}\right)  ,
 \label{Vud}
\end{eqnarray}
where $d_{d}$  and $d_{ud}$  are the respective interaction
functions in first quantization. The elastic collisions
between atoms and molecules are described by the interaction
term
\begin{eqnarray}
\hat{V}_{el}&=&\int d^{3}r \biggl\lbrack  {U{ } _{a}\over 2}\hat{\psi
 }^{+}_{a}\left( {\bf r}\right) \hat{\psi }^{+}_{a}\left( {\bf
 r}\right) \hat{\psi }_{a}\left( {\bf r}\right) \hat{\psi }_{a}\left(
 {\bf r}\right) +{U{ } _{m}\over 2}\sum\limits^{}_{\alpha ,\alpha
 ^\prime }\hat{\psi }^{+}_{\alpha }\left( {\bf r}\right) \hat{\psi
 }^{+}_{\alpha ^\prime }\left( {\bf r}\right) \hat{\psi }_{\alpha
 ^\prime }\left( {\bf r}\right) \hat{\psi }_{\alpha }\left( {\bf
 r}\right)  \nonumber
\\
&&+U_{am}\sum\limits^{}_{\alpha }\hat{\psi }^{+}_{a}\left( {\bf
 r}\right) \hat{\psi }^{+}_{\alpha }\left( {\bf r}\right) \hat{\psi
 }_{\alpha }\left( {\bf r}\right) \hat{\psi }_{a}\left( {\bf r}\right
) \biggr\rbrack  ,\qquad \alpha ,\alpha '{}=m,u,d
\end{eqnarray}
including terms proportional to the zero-momentum atom-atom,
molecule-molecule, and atom-molecule elastic scattering lengths
($a_{a}$,
$a_{m}$, and $a_{am}$, respectively),
\begin{equation}
U_{a}={4\pi \over m}a_{a},\quad U_{m}={2\pi \over m}a_{m},\quad
 U_{am}={3\pi \over m}a_{am} .
\end{equation}
The different numerical factors in the numerators
reflect the different reduced masses.

Finally, the total Hamiltonian of the atom-molecule gas
is represented as the sum of free field Hamiltonians and
interactions related to all collision processes taken into
account,
\begin{equation}
\hat{H}=\hat{H}_{\text{free}}+\hat{V}_{h}+\hat{V}^{\dag }_{h}
+\sum\limits^{}_{d}\left( \hat{H}_{d}+\hat{V}_{d}+\hat{V}^{\dag
 }_{d}\right) +\sum\limits^{}_{u}\hat{H}_{u}
+\sum\limits^{}_{u,d}\left( \hat{V}_{ud}+\hat{V}^{\dag }_{ud}\right)
+\hat{V}_{el}.
\end{equation}
The Hamiltonian was described above in the Schr\"odinger
picture  using the coordinate representation. The momentum
representation is more suitable for homogeneous systems. It
can also be used for an approximate analysis of
inhomogeneous systems in the local density approximation.
The operator equations of motion in the Heisenberg picture
will be used below for the analysis of non-mean-field
effects. The Heisenberg field operator in the momentum
representation $\hat{\Psi }_{\alpha }\left( {\bf p},t\right) $ is
 related to the corresponding
Schr\"odinger operator in the coordinate representation $\hat{\psi
 }_{\alpha }\left( {\bf r}\right) $
as
\begin{equation}
\hat{\Psi }_{\alpha }\left( {\bf p},t\right) =\left( 2\pi \right)
 ^{-3/2}\int d^{3}r e^{-i{\bf p}{\bf r}}\hat{\Omega }^{-1}\hat{\psi
 }_{\alpha }\left( {\bf r}\right) \hat{\Omega } ,
\end{equation}
where the time evolution operator $\hat{\Omega }$ satisfies the
equation
\begin{equation}
i{\partial \over \partial t}\hat{\Omega }=\hat{H} \hat{\Omega } .
\end{equation}
As a result, the transformed Hamiltonian can be
expressed in terms of the field operators in the momentum
representation as
\begin{eqnarray}
\hat{\tilde{H}}&=&\hat{\Omega }^{-1}\hat{H}\hat{\Omega }=\int
 d^{3}p\Biggl\{\left\lbrack {p{ } ^{2}\over 2m}+\epsilon _{a}\left(
 t\right) \right\rbrack \hat{\Psi }^{\dag }_{a}\left( {\bf p},t\right
) \hat{\Psi }_{a}\left( {\bf p},t\right)  \nonumber
\\
&&+{p{ } ^{2}\over 4m}\hat{\Psi }^{\dag }_{m}\left( {\bf p},t\right)
 \hat{\Psi }_{m}\left( {\bf p},t\right) + \sum\limits^{}_{\alpha
 =u,d}\left\lbrack  {p{ } ^{2}\over 4m} -E_{\alpha }\right\rbrack
 \hat{\Psi }^{\dag }_{\alpha }\left( {\bf p},t\right) \hat{\Psi
 }_{\alpha }\left( {\bf p},t\right) \Biggr\}+ \nonumber
\\
&&+\hat{\tilde{V}}_{h}+\hat{\tilde{V}}^{\dag }_{h}+
 \sum\limits^{}_{d}\left( \hat{\tilde{V}}_{d}+\hat{\tilde{V}}^{\dag
 }_{d}\right) +\sum\limits^{}_{u,d}\left( \hat{\tilde{V}}_{ud}
+\hat{\tilde{V}}^{\dag }_{ud}\right)  , \label{genH}
\end{eqnarray}
where the interaction operators are represented as
\begin{eqnarray}
\hat{\tilde{V}}_{h}=\left( 2\pi \right) ^{-3/2}&\int &d^{3}p
 d^{3}p^\prime \tilde{d}_{h}\left( {\bf p}-{\bf p}^\prime \right)
 \hat{\Psi }^{\dag }_{m}\left( {\bf p}+{\bf p}^\prime ,t\right)
 \hat{\Psi }_{a}\left( {\bf p},t\right) \hat{\Psi }_{a}\left( {\bf
 p}^\prime ,t\right)  \label{Vhm}
\\
\hat{\tilde{V}}_{d}=\left( 2\pi \right) ^{-3/2}&\int &d^{3}p^\prime
 d^{3}p_{d}d^{3}p_{m}\tilde{d}_{d}\left( {\bf p}_{d}-2{\bf p}^\prime
 \right) \hat{\Psi }^{\dag }_{a}\left( {\bf p}^\prime ,t\right)
 \hat{\Psi }^{\dag }_{d}\left( {\bf p}_{d},t\right)  \nonumber
\\
&&\times \hat{\Psi }_{m}\left( {\bf p}_{m},t\right) \hat{\Psi
 }_{a}\left( {\bf p}^\prime +{\bf p}_{d}-{\bf p}_{m},t\right)
\\
\hat{\tilde{V}}_{ud}=\left( 2\pi \right) ^{-3/2}&\int
 &d^{3}p_{m}d^{3}p_{d}d^{3}p_{u}\tilde{d}_{ud}\left( {\bf p}_{u}-{\bf
 p}_{d}\right) \hat{\Psi }^{\dag }_{u}\left( {\bf p}_{u},t\right)
 \hat{\Psi }^{\dag }_{d}\left( {\bf p}_{d},t\right)  \nonumber
\\
&&\times \hat{\Psi }_{m}\left( {\bf p}_{m},t\right) \hat{\Psi
 }_{m}\left( {\bf p}_{u}+{\bf p}_{d}-{\bf p}_{m},t\right)  ,
\end{eqnarray}
with
\begin{eqnarray}
\tilde{d}_{h}\left( p\right) =\int d^{3}\rho V_{h}\left( \rho \right)
 \exp\left( -i{\bf p}\bm{\rho}/2\right)  \label{dhm}
\\
\tilde{d}_{d}\left( p\right) =\left( 2\pi \right) ^{-3/2}\int
 d^{3}\rho d_{d}\left( \rho \right) \exp\left( -i{\bf p}\bm{\rho}
/3\right)  \label{ddm}
\\
\tilde{d}_{ud}\left( p\right) =\left( 2\pi \right) ^{-3/2}\int
 d^{3}\rho d_{ud}\left( \rho \right) \exp\left( -i{\bf p}\bm{\rho}
/2\right)  . \label{dudm}
\end{eqnarray}
The transformation of the elastic collision terms is not
presented here as the latter can be neglected in the analysis of
non-mean-field effects below, carried out in the momentum
representation.

\section{Mean field approximation\label{SecMeanField}}

\subsection{Trial function}

Mean field equations can be derived by a variational
method (see Refs.\ \cite{BR86,EDCB96}) using a trial function
in the form of a coherent state. This approach is applicable
to the description of macroscopically occupied states, such
as the atomic and resonant molecular states in the system
considered here. However, ``dump'' states --- ``hot''
products of the deactivating reactions  (\ref{AMCol}) and
(\ref{MMCol}) --- are formed in a wide energy spectrum,
escape the trap very fast, and therefore have low occupation.
The form of Eqs.\ (\ref{Vd}) and (\ref{Vud}) demonstrates
that the resulting atoms and molecules are formed in these
reactions as correlated pairs. These reasons suggest the
following choice of a trial function (see Ref.\
\cite{YBJW00}),
\begin{eqnarray}
|\Phi \rangle =&&\Bigl\lbrack 1+\int d^{3}r
 d^{3}r_{m}\sum\limits^{}_{d}\varphi _{d}\left( {\bf r},{\bf
 r}_{m},t\right) \hat{\psi }^{+}_{a}\left( {\bf r}\right) \hat{\psi
 }^{+}_{d}\left( {\bf r}_{m}\right)  \nonumber
\\
&&+\int d^{3}r_{1}d^{3}r_{2}\sum\limits^{}_{u,d}\varphi _{ud}\left(
 {\bf r}_{1},{\bf r}_{2},t\right) \hat{\psi }^{+}_{u}\left( {\bf
 r}_{1}\right) \hat{\psi }^{+}_{d}\left( {\bf r}_{2}\right)
 \Bigr\rbrack  |\varphi _{0},\varphi _{m}\rangle  . \label{TrFun}
\end{eqnarray}
It contains a coherent state
\begin{equation}
|\varphi _{0},\varphi _{m}\rangle =\exp\Bigl\{ \int
 d^{3}r\bigl\lbrack \varphi _{0}\left( {\bf r},t\right) \hat{\psi }^{
+}_{a}\left( {\bf r}\right) +\varphi _{m}\left( {\bf r},t\right)
 \hat{\psi }^{+}_{m}\left( {\bf r}\right) \bigr\rbrack \Bigr\}
|0\rangle  ,
\end{equation}
formed by a product of exponential operators, involving
atomic ($\varphi _{0}$) and molecular ($\varphi _{m}$) condensate
 states, and
operating on the vacuum state $|0\rangle $. The linear factor
preceding it in Eq.\ (\ref{TrFun}) includes the fields
$\varphi _{d}\left( {\bf r},{\bf r}_{2},t\right) $ and $\varphi
 _{ud}\left( {\bf r}_{1},{\bf r}_{2},t\right) $, which are the
 correlated dump
states of the products in reactions (\ref{AMCol}) and
(\ref{MMCol}), respectively. Another constraint, that
follows from the large energy difference between the dump
states and the resonant state, is the condition
\begin{equation}
\int d^{3}r\varphi ^{*}_{0}\left( {\bf r},t\right) \varphi _{d}\left(
 {\bf r},{\bf r}_{m},t\right) =0 \label{ortphi}
\end{equation}
[see discussion following Eq.\ (\ref{phidkro}) for
justification].

The trial wavefunction (\ref{TrFun}) can now be
substituted into a variational functional (see Refs.\
\cite{BR86,EDCB96})
\begin{equation}
\int\limits^{\infty }_{-\infty }dt{\langle \Phi |i {\partial \over
 \partial t}-\hat{H}|\Phi \rangle \over \langle \Phi |\Phi \rangle } .
\end{equation}
In the mean-field calculations one can use the
approximation of zero-range interaction $V_{h}\left( {\bf r}-{\bf
 r}^\prime \right) =g\delta \left( {\bf r}-{\bf r}^\prime \right) $,
and represent the atom-molecule coupling $\hat{V}_{h}$  [see Eq.\
(\ref{Vh})] in the simpler form
\begin{equation}
\hat{V}_{h}=g\int d^{3}r \hat{\psi }^{+}_{m}\left( {\bf r}\right)
 \hat{\psi }_{a}\left( {\bf r}\right) \hat{\psi }_{a}\left( {\bf
 r}\right)  . \label{VH}
\end{equation}
The parameter $g$ is related to integral of $V_{h}\left( \rho \right)
 $  and to
the peak value of $\tilde{d}_{h}\left( p\right) $ [see Eq.\
(\ref{dhm})] as
\begin{equation}
g=\int d^{3}\rho V_{h}\left( \rho \right) =\tilde{d}_{h}\left(
 0\right)  . \label{gVd}
\end{equation}
The interactions responsible for the deactivating collisions
(\ref{AMCol}) and (\ref{MMCol}) are kept as finite-range functions of
the distance between the reaction products [see Eqs.\ (\ref{Vd}) and
(\ref{Vud})].

Neglecting terms of the order of $\varphi ^{3}_{d}$  and $\varphi
 ^{3}_{u}$, and taking
into account Eq.\ (\ref{ortphi}), the use of a standard
variational procedure (see Ref.\ \cite{EDCB96}) then leads to
a set of coupled equations for the atomic ($\varphi _{0}$) and
 molecular
($\varphi _{m}$) condensate fields (or ``wavefunctions''), as well as
 for
the dump states ($\varphi _{d}$  and $\varphi _{ud}$):
 \begin{subequations}
\begin{eqnarray}
i\dot{\varphi }_{0}\left( {\bf r},t\right) =&&\left( \hat{H}_{a}
+U_{a}|\varphi _{0}\left( {\bf r},t\right) |^{2}+U_{am}|\varphi
 _{m}\left( {\bf r},t\right) |^{2}\right) \varphi _{0}\left( {\bf
 r},t\right)  \nonumber
\\
&&+2g^{*}\varphi ^{*}_{0}\left( {\bf r},t\right) \varphi _{m}\left(
 {\bf r},t\right) +Q\left( {\bf r},t\right) \varphi ^{*}_{m}\left(
 {\bf r},t\right)  \label{phi0d}
\\
i\dot{\varphi }_{m}\left( {\bf r},t\right) =&&\left( \hat{H}_{m}
+U_{am}|\varphi _{0}\left( {\bf r},t\right) |^{2}+U_{m}|\varphi
 _{m}\left( {\bf r},t\right) |^{2}\right) \varphi _{m}\left( {\bf
 r},t\right)  \nonumber
\\
&&+g\varphi ^{2}_{0}\left( {\bf r},t\right) +Q\left( {\bf r},t\right)
 \varphi ^{*}_{0}\left( {\bf r},t\right) +Q_{m}\left( {\bf r},t\right
) \varphi ^{*}_{m}\left( {\bf r},t\right)  \label{phimd}
\\
i\dot{\varphi }_{d}\left( {\bf r}_{1},{\bf r}_{2},t\right) =&&\left(
 {1\over 2m}\hat{{\bf p}}^{2}_{1}+{1\over 4m}\hat{{\bf
 p}}^{2}_{2}-E_{d}\right) \varphi _{d}\left( {\bf r}_{1},{\bf
 r}_{2},t\right)  \nonumber
\\
&&+d_{d}\left( |{\bf r}_{1}-{\bf r}_{2}|\right) \varphi _{0}\left(
 {{\bf r}_{1}+2{\bf r}{ } _{2}\over 3},t\right) \varphi _{m}\left(
 {{\bf r}_{1}+2{\bf r}{ } _{2}\over 3},t\right)  \label{phid}
\\
i\dot{\varphi }_{ud}\left( {\bf r}_{1},{\bf r}_{2},t\right) =&&\left(
 {1\over 4m}\hat{{\bf p}}^{2}_{1}+{1\over 4m}\hat{{\bf
 p}}^{2}_{2}-E_{u}-E_{d}\right) \varphi _{ud}\left( {\bf r}_{1},{\bf
 r}_{2},t\right)  \nonumber
\\
&&+d_{ud}\left( |{\bf r}_{1}-{\bf r}_{2}|\right) \varphi
 ^{2}_{m}\left( {{\bf r}_{1}+{\bf r}{ } _{2}\over 2},t\right)  ,
 \label{phiud}
\end{eqnarray}
\end{subequations} where
\begin{eqnarray}
Q\left( {\bf r},t\right) &=&\int
 d^{3}r_{1}d^{3}r_{2}\sum\limits^{}_{d}d^{*}_{d}\left( |{\bf
 r}_{1}-{\bf r}_{2}|\right) \varphi _{d}\left( {\bf r}_{1},{\bf
 r}_{2},t\right) \delta \left( {\bf r}-{{\bf r}_{1}+2{\bf r}{ }
 _{2}\over 3}\right)  \label{Q}
\\
Q_{m}\left( {\bf r},t\right) &=&2\int
 d^{3}r_{1}d^{3}r_{2}\sum\limits^{}_{u,d}d^{*}_{ud}\left( |{\bf
 r}_{1}-{\bf r}_{2}|\right) \varphi ^{*}_{m}\left( {\bf r},t\right)
 \varphi _{ud}\left( {\bf r}_{1},{\bf r}_{2},t\right) \times \delta
 \left( {\bf r}-{{\bf r}_{1}+{\bf r}{ } _{2}\over 2}\right)  .
 \nonumber
\\
&& \label{Qm}
\end{eqnarray}
\subsection{Dump state elimination}

The procedure used in Ref.\ \cite{YBJW00} to eliminate
the dump states is similar to the Weisskopf-Wigner method in
the theory of spontaneous emission (see Ref.\ \cite{Agarwal}).
Equation (\ref{phid}) is of the form of a Schr\"odinger equation
for two free particles with a source (the last term in the
right-hand side). Such an equation can be solved by applying
the Green's function method for free particles, with the
result
\begin{eqnarray}
&&\varphi _{d}\left( {\bf r}_{1},{\bf r}_{2},t\right) =-{i\over \left
( 2\pi \right) { } ^{6}}\int\limits^{t}_{-\infty }dt^\prime \int
 d^{3}P d^{3}p d^{3}\rho  d^{3}R d_{d}\left( \rho \right) \varphi
 _{m}\left( {\bf R},t^\prime \right) \varphi _{0}\left( {\bf
 R},t^\prime \right) e^{-i{\bf p}\bm{\rho}-i{\bf P}{\bf R}} \nonumber
\\
&&\times \exp\left\lbrack -i\left( {P{ } ^{2}\over 6m}+{3p{ }
 ^{2}\over 4m}-E_{d}-i0\right) \left( t-t^\prime \right) +i{\bf
 P}{{\bf r}_{1}+2{\bf r}{ } _{2}\over 3}+i{\bf p}\left( {\bf
 r}_{1}-{\bf r}_{2}\right) \right\rbrack  . \label{phidGf}
\end{eqnarray}
Here ${\bf R}$ is the center-of-mass position of the three-atom
system, $\bm{\rho}$ is the radius vector of the reaction
products, and ${\bf P}$, ${\bf p}$ are the corresponding momenta.
 Since $\varphi _{m}\left( {\bf R},t\right) $
and $\varphi _{0}\left( {\bf R},t\right) $ are condensate
 wavefunctions, the Fourier transform
of their product [the integral over ${\bf R}$ in Eq.\ (\ref{phidGf})]
vanishes if $P>1/b$, where $b\sim \left( m\omega _{\text{trap}}\right
) ^{-1/2}$  is a characteristic
size of the condensate and $\omega _{\text{trap}}$  is the trap
 frequency.
Therefore $P^{2}/\left( 6m\right) <\omega _{\text{trap}}$  is
 negligible compared to $E_{d}\gg \omega _{\text{trap}}$.
This fact allows us also to neglect the time dependence of
$\varphi _{m}\left( {\bf R},t\right) $ and $\varphi _{0}\left( {\bf
 R},t\right) $ in the integration over $t^\prime $,  and thus
obtain the simplified expression
\begin{eqnarray}
\varphi _{d}\left( {\bf r}_{1},{\bf r}_{2},t\right) =&&-{1\over \left
( 2\pi \right) { } ^{3}}\varphi _{0}\left( {{\bf r}_{1}+2{\bf r}{ }
 _{2}\over 3},t\right) \varphi _{m}\left( {{\bf r}_{1}+2{\bf r}{ }
 _{2}\over 3},t\right)  \nonumber
\\
&&\times \int d^{3}p d^{3}\rho  {d_{d}\left( \rho \right)
 \exp\left\lbrack i{\bf p}\left( {\bf r}_{2}-{\bf
 r}_{1}-\bm{\rho}\right) \right\rbrack \over 3p^{2}/\left( 4m\right)
 -E_{d}-i0} . \label{phidkro}
\end{eqnarray}
The atom-molecule pair is thus formed with a momentum
of relative motion
\begin{equation}
P_{d}=\sqrt{4mE_{d}/3} . \label{Pd}
\end{equation}
The function $\varphi _{d}\left( {\bf r}_{1},{\bf r}_{2},t\right) $
 is a rapidly oscillating
function of the coordinates, and therefore condition
(\ref{ortphi}) is justified.

Substituting Eqs.\ (\ref{phidkro}) and (\ref{ddm}) into
Eqs.\ (\ref{Q}) one obtains
\begin{equation}
Q\left( {\bf r},t\right) =-\left( \delta _{a}+ik_{a}/2\right) \varphi
 _{0}\left( {\bf r},t\right) \varphi _{m}\left( {\bf r},t\right)  ,
 \label{QDelGam}
\end{equation}
where
\begin{equation}
\delta _{a}+i{k{ } _{a}\over 2}=\sum\limits^{}_{d} \int d^{3}p {
|\tilde{d}_{d}\left( 3p\right) |{ } ^{2}\over 3p^{2}/\left( 4m\right)
 -E_{d}-i0} .
\end{equation}
Using the well-known identity $\left( x-i0\right) ^{-1}={\cal
 P}x^{-1}+i\pi \delta \left( x\right) $,
where ${\cal P}$ denotes the Cauchy principal part of the integral,
allows us to obtain explicit expressions for $k_{a}$  and $\delta
 _{a}$,
\begin{eqnarray}
k_{a}={4\pi ^{2}m\over 3} \sum\limits^{}_{d}P_{d}\left
|\tilde{d}_{d}\left( 3P_{d}\right) \right|^{2} \label{gamma}
\\
\delta _{a}={16\pi m\over 3}\sum\limits^{}_{d}{\cal
 P}\int\limits^{\infty }_{0}dp{p^{2}|\tilde{d}_{d}\left( 3p\right) |{
 } ^{2}\over p^{2}-P{ } ^{2}_{d}} . \label{delta}
\end{eqnarray}
A similar analysis, starting from Eq.\ (\ref{phiud}), gives
\begin{equation}
Q_{m}\left( {\bf r},t\right) =-\left( \delta _{m}+ik_{m}\right)
 \varphi ^{2}_{m}\left( {\bf r},t\right)  , \label{QmDelGam}
\end{equation}
where
\begin{eqnarray}
k_{m}=8\pi ^{2}m\sum\limits^{}_{u,d}P_{ud}\left|\tilde{d}_{ud}\left(
 2P_{ud}\right) \right|^{2} \label{gammam}
\\
\delta _{m}=16\pi m\sum\limits^{}_{u,d}{\cal P}\int\limits^{\infty
 }_{0}dp{p^{2}|\tilde{d}_{ud}\left( 2p\right) |{ } ^{2}\over p^{2}-P{
 } ^{2}_{ud}} , \label{deltam}
\end{eqnarray}
$P_{ud}=\sqrt{2m\left( E_{u}+E_{d}\right) }$, and $\tilde{d}_{ud}$
 is defined by Eq.\ (\ref{dudm}).
Substituting Eqs.\ (\ref{QDelGam}) and (\ref{QmDelGam}) into
Eqs.\ (\ref{phi0d}) and (\ref{phimd}) one finally obtains a pair
of coupled Gross-Pitaevskii equations (see Ref.\ \cite{GP})
\begin{subequations} \label{GPE}
\begin{eqnarray}
i\dot{\varphi }_{0}=\left( \hat{H}_{a}+U_{a}|\varphi _{0}|^{2}+U_{am}
|\varphi _{m}|^{2}\right) \varphi _{0}+2g^{*}\varphi ^{*}_{0}\varphi
 _{m}-\left( \delta _{a}+ik_{a}/2\right) |\varphi _{m}|^{2}\varphi
 _{0} \label{GPEa}
\\
i\dot{\varphi }_{m}=\left( \hat{H}_{m}+U_{am}|\varphi _{0}|^{2}+U_{m}
|\varphi _{m}|^{2}\right) \varphi _{m}+g\varphi ^{2}_{0} \nonumber
\\
-\left\lbrack \left( \delta _{a}+ik_{a}/2\right) |\varphi _{0}|^{2}
+\left( \delta _{m}+ik_{m}\right) |\varphi _{m}|^{2}\right\rbrack
 \varphi _{m} . \label{GPEm}
\end{eqnarray}
\end{subequations} The parameters $\delta _{a}$, $k_{a}$, $\delta
 _{m}$, and $k_{m}$, which
are expressed in terms of $\tilde{d}_{d}$  and $\tilde{d}_{ud}$  [see
 Eqs.\ (\ref{gamma}),
(\ref{delta}), (\ref{gammam}), and (\ref{deltam})], describe the
shift and the width of the resonance due to the deactivating
collisions with atoms and molecules, respectively. The
parameters $k_{a}$  and $k_{m}$  are the corresponding rate
 constants. The
shifts $\delta _{a}$  and $\delta _{m}$  can be incorporated in the
 interactions $U_{am}$
and $U_{m}$, respectively.

In the case of a time-independent magnetic field and
large resonant detuning, and neglecting the decay described
by the imaginary terms, Eqs.\ (\ref{GPEa}) and (\ref{GPEm})
can be reduced to a single Gross-Pitaevskii equation with an
effective scattering length $a_{a}\left\lbrack 1-\Delta /\left(
 B-B_{0}\right) \right\rbrack $. The
phenomenological resonance strength $\Delta $ can be measured in
experiments or obtained from two-body scattering
calculations. It is related to the atom-molecule coupling
constant $g$ of Eq.\ (\ref{VH}) as
\begin{equation}
|g|^{2}=2\pi |a_{a}|\mu \Delta /m , \label{Delta}
\end{equation}
from which the value of $g$ can be deduced, given $\Delta $.

The atom-molecule coupling terms in Eqs.\ (\ref{GPEa}) and
(\ref{GPEm}) are proportional to the atomic mean field $\varphi
 _{0}$. Therefore the
dissociation of molecules into the atomic vacuum ($\varphi _{0}=0$)
 is forbidden. In
the absence of molecule-molecule collisions ($k_{m}=0$) a pure
 molecular BEC
is a stationary solution of the mean-field equations. This solution
 is,
nevertheless, unstable, and a proper description of the dissociation
requires the use of a non-mean-field theory, taking into account
 atomic
field fluctuations (see Sec.\ \ref{SecParAppr} below).

In experiments with a trapped BEC the kinetic energy
terms in Eqs.\ (\ref{GPEa}) and (\ref{GPEm}), arising from
$\hat{H}_{a}$  and $\hat{H}_{m}$  [see Eq.\ (\ref{HaHm})], can be
 generally
neglected according to the Thomas-Fermi approximation (see
Ref.\ \cite{TF}). In this case $\varphi _{0}\left( {\bf r},t\right) $
 and $\varphi _{m}\left( {\bf r},t\right) $ depend on
${\bf r}$ only parametrically and Eqs.\ (\ref{GPEa}) and
(\ref{GPEm}) are reduced to a set of ordinary differential
equations in $t$. A different situation takes place in the
case of an expanding BEC considered below in Sec.\
\ref{SecExpan}.

When the kinetic energy terms are neglected, the system can be
described by a set of real equations, similar to the optical Bloch
equations (see Ref.\ \cite{YBJW00}) for the new real variables, the
atomic and molecular ``populations'' (condensate densities)
\begin{equation}
n_{0}\left( {\bf r},t\right) =|\varphi _{0}\left( {\bf r},t\right)
|^{2}, \quad n_{m}\left( {\bf r},t\right) =|\varphi _{m}\left( {\bf
 r},t\right) |^{2},
\end{equation}
and ``coherencies''
\begin{equation}
u\left( {\bf r},t\right) =2\text{Re}\left( g\varphi ^{2}_{0}\left(
 {\bf r},t\right) \varphi ^{*}_{m}\left( {\bf r},t\right) \right)
 ,\qquad v\left( {\bf r},t\right) =-2\text{Im}\left( g\varphi
 ^{2}_{0}\left( {\bf r},t\right) \varphi ^{*}_{m}\left( {\bf
 r},t\right) \right)  .
\end{equation}
Equations (\ref{GPEa}) and (\ref{GPEm}) lead to the following
equations of motion for the new variables,  \begin{subequations}
\label{deneq}
\begin{eqnarray}
\dot{n}_{0}\left( {\bf r},t\right) &=&2v\left( {\bf r},t\right)
 -2\Gamma _{a}\left( {\bf r},t\right) n_{0}\left( {\bf r},t\right)
 \label{deneqn}
\\
\dot{n}_{m}\left( {\bf r},t\right) &=&-v\left( {\bf r},t\right)
 -2\Gamma _{m}\left( {\bf r},t\right) n_{m}\left( {\bf r},t\right)
\\
\dot{v}\left( {\bf r},t\right) &=&D\left( {\bf r},t\right) u\left(
 {\bf r},t\right) -\left\lbrack 2\Gamma _{a}\left( {\bf r},t\right)
+\Gamma _{m}\left( {\bf r},t\right) \right\rbrack v\left( {\bf
 r},t\right)  \nonumber
\\
&&+2|g|^{2}n_{0}\left( {\bf r},t\right) \left\lbrack 4n_{m}\left(
 {\bf r},t\right) -n_{0}\left( {\bf r},t\right) \right\rbrack
\\
\dot{u}\left( {\bf r},t\right) &=&-D\left( {\bf r},t\right) v\left(
 {\bf r},t\right) -\left\lbrack 2\Gamma _{a}\left( {\bf r},t\right)
+\Gamma _{m}\left( {\bf r},t\right) \right\rbrack u\left( {\bf
 r},t\right)  .
\end{eqnarray}
 \end{subequations} Here
\begin{eqnarray}
D\left( {\bf r},t\right) =2\epsilon _{a}\left( t\right) +V\left(
 r\right) +\left( 2U_{a}-U_{am}+\delta _{a}\right) n_{0}\left( {\bf
 r},t\right)  \nonumber
\\
+\left( 2U_{am}-2\delta _{a}-U_{m}+\delta _{m}\right) n_{m}\left(
 {\bf r},t\right)  \label{D}
\\
V\left( {\bf r}\right) =2V_{a}\left( {\bf r}\right) -V_{m}\left( {\bf
 r}\right) , \nonumber
\end{eqnarray}
and
\begin{equation}
\Gamma _{a}\left( {\bf r},t\right) ={1\over 2}k_{a}n_{m}\left( {\bf
 r},t\right)  ,\qquad \Gamma _{m}\left( {\bf r},t\right) ={1\over
 2}k_{a}n_{0}\left( {\bf r},t\right) +k_{m}n_{m}\left( {\bf
 r},t\right)  . \label{Gammaam}
\end{equation}
A form of such equations similar to the vector form of the
optical Bloch equations is derived in Ref.\ \cite{VYA01}.

\subsection{Expanding Bose-Einstein condensate\label{SecExpan}}

Consider an atomic BEC confined in a harmonic trap
potential
\begin{equation}
V_{a}\left( {\bf r}\right) ={m\over 2}\sum\limits^{3}_{j=1}\omega
 ^{2}_{j}r^{2}_{j} .
\end{equation}
In the Thomas-Fermi regime, when the kinetic energy is negligible
compared to the chemical potential $\epsilon _{0}$, the atomic mean
 field can be
described by the stationary Thomas-Fermi solution (see Ref.\
 \cite{TF})
\begin{equation}
\varphi _{TF}\left( {\bf r}\right) =N^{-1/2}_{0}\left(
 n_{\text{peak}}-{m\over 2U{ } _{a}} \sum\limits^{3}_{j=1}\omega
 ^{2}_{j}r^{2}_{j}\right) ^{1/2}, n_{\text{peak}}=\left(
 {15m^{3}\omega _{1}\omega _{2}\omega _{3}N{ } _{0}\over 128\pi
 ^{2}\sqrt{2\pi }}\right) ^{2/5}a^{-3/5}_{a}, \label{ThomasFermi}
\end{equation}
where $N_{0}$  is the initial number of atoms, $n_{\text{peak}}$  is
 the peak
density, and $\epsilon _{0}=U_{a}n_{\text{peak}}$.

An expanding BEC is formed when the atomic trap containing the BEC
is turned off. The following expansion of pure atomic BEC has been
considered in Refs.\ \cite{expan}. In this case a solution of the
 single
Gross-Pitaevskii equation
\begin{equation}
i\dot{\varphi }_{0}\left( {\bf r},t\right) =\left\lbrack -{1\over
 2m}\nabla ^{2}+U_{a}|\varphi _{0}\left( {\bf r},t\right)
|^{2}\right\rbrack \varphi _{0}\left( {\bf r},t\right)  \label{GPa1}
\end{equation}
can be represented in the form
\begin{equation}
\varphi _{0}\left( {\bf r},t\right) =A\left( t\right) \Phi _{0}\left(
 \bm{\rho },t\right) e^{i}{ } ^{S} , \label{phia}
\end{equation}
using the scaled coordinates
\begin{equation}
\rho _{j}=r_{j}/b_{j}\left( t\right) , 1\le j\le 3,
\end{equation}
a scaling factor
\begin{equation}
A\left( t\right) =\left( b_{1}\left( t\right) b_{2}\left( t\right)
 b_{3}\left( t\right) \right) ^{-1/2},
\end{equation}
and the phase
\begin{equation}
S\left( {\bf r},t\right)
 =m\sum\limits^{3}_{j=1}r^{2}_{j}{\dot{b}_{j}\left( t\right) \over
 2b_{j}\left( t\right) }+S_{0}\left( t\right)
\end{equation}
which incorporates most of the contribution of the kinetic
energy.

The initial conditions at the start of the expansion $t=t_{\exp}$  are
$b_{j}\left( t_{\exp}\right) =1$, $\dot{b}_{j}\left( t_{\exp}\right)
 =0$, $S_{0}\left( t_{\exp}\right) =0$, and
$\Phi _{0}\left( \bm{\rho },t_{\exp}\right) =\varphi _{0}\left(
 \bm{\rho },t_{\exp}\right) =\varphi _{TF}\left( \bm{\rho }\right) $.
 Substitution of Eq.\
(\ref{phia}) into Eq.\ (\ref{GPa1}) leads to the following equation
 for the
transformed mean field $\Phi _{0}\left( \bm{\rho },t\right) $
\begin{equation}
i \dot{\Phi }_{0}\left( \bm{\rho },t\right) =\left\lbrack -{1\over
 2m}\sum\limits^{3}_{j=1}{1\over b{ } ^{2}_{j}} {\partial { }
 ^{2}\over \partial \rho { } ^{2}_{j}}+\dot{S}_{0}+{m\over 2}
 \sum\limits^{3}_{j=1}\ddot{b}_{j}b_{j}\rho ^{2}_{j}+A^{2}\left(
 t\right) U_{a}|\Phi _{0}\left( \bm{\rho },t\right)
|^{2}\right\rbrack \Phi _{0}\left( \bm{\rho },t\right) . \label{GPa2}
\end{equation}
Following Ref.\ \cite{expan} let us take the scales $b_{j}$  as
 solutions
of the set of equations
\begin{equation}
\ddot{b}_{j}\left( t\right) =\omega ^{2}_{j}A^{2}\left( t\right)
/b_{j}\left( t\right)  \label{exp_b}
\end{equation}
and $\dot{S}_{0}\left( t\right) =-\epsilon _{0}A^{2}\left( t\right)
 $. As has been shown in Ref.\ \cite{expan} the
residual kinetic energy terms can be neglected in the Thomas-Fermi
 regime
and Eq.\ (\ref{GPa2}) is satisfied by the stationary Thomas-Fermi
 solution
$\Phi _{0}\left( \bm{\rho },t\right) =\varphi _{TF}\left( \bm{\rho
 }\right) $.

Solutions of Eq.\ (\ref{exp_b}) (see Ref.\ \cite{expan}) demonstrate
that the expansion is ballistic after acceleration during a time
 interval
$t\sim \min\left( \omega ^{-1}_{j}\right) $. In experiments the
 expansion is started at a large detuning,
when the molecular occupation is negligibly small. The resonance is
approached and the molecules are formed when the density is
 substantially
reduced. Therefore the terms proportional to $U_{m}$, $U_{am}$,
 $\delta _{a}$, $\delta _{m}$, and $V_{m}\left( {\bf r}\right) $  in
Eq.\ (\ref{GPE}) can be neglected. Since  the molecules are formed at
 the
ballistic stage of the expansion and inherit the velocity of the
 atoms they
are formed from, the molecular field can be represented in the form
\begin{equation}
\varphi _{m}\left( {\bf r},t\right) =A\left( t\right) \Phi _{m}\left(
 \bm{\rho },t\right) e^{2i}{ } ^{S} . \label{phim}
\end{equation}
Substitution of Eqs.\ (\ref{phia}) and (\ref{phim}) into Eq.\
(\ref{GPE}) leads to the following set of coupled equations for the
transformed mean fields (see Ref.\ \cite{YB04}):
\begin{eqnarray}
i\dot{\Phi }_{0}\left( \bm{\rho },t\right) &=&\left\lbrack \epsilon
 _{a}\left( t\right) -{i\over 2}A^{2}\left( t\right) k_{a}|\Phi
 _{m}\left( \bm{\rho },t\right) |^{2}\right\rbrack \Phi _{0}\left(
 \bm{\rho },t\right) +2A\left( t\right) g^{*}\Phi ^{*}_{0}\left(
 \bm{\rho },t\right) \Phi _{m}\left( \bm{\rho },t\right)  \nonumber
\\
&&+A^{2}\left( t\right) \left( U_{a}|\Phi _{0}\left( \bm{\rho
 },t\right) |^{2}-\epsilon _{0}+{m\over 2}
 \sum\limits^{3}_{j=1}\omega ^{2}_{j}\rho ^{2}_{j}\right) \Phi
 _{0}\left( \bm{\rho },t\right)  \label{GPexpa}
\\
i\dot{\Phi }_{m}\left( \bm{\rho },t\right) &=&-iA^{2}\left( t\right)
 \biggl\lbrack {1\over 2}k_{a}|\Phi _{0}\left( \bm{\rho },t\right)
|^{2}+k_{m}|\Phi _{m}\left( \bm{\rho },t\right) |^{2}\biggr\rbrack
 \Phi _{m}\left( \bm{\rho },t\right) +A\left( t\right) g\Phi
 ^{2}_{0}\left( \bm{\rho },t\right)  \nonumber
\\
&&-A^{2}\left( t\right) \left( \epsilon
 _{0}-m\sum\limits^{3}_{j=1}\omega ^{2}_{j}\rho ^{2}_{j}\right) \Phi
 _{m}\left( \bm{\rho },t\right)  \label{GPexpm}
\end{eqnarray}
The residual kinetic energy terms can be neglected in the Thomas-Fermi
regime as well as in the case of a pure atomic BEC. The loss
 processes and
atom-molecule transitions distort $\Phi _{0}\left( \bm{\rho },t\right
) $ from the Thomas-Fermi
shapes, leading to additional energy shift compared to the pure
 atomic case
[the last term in Eq.\ (\ref{GPexpa})]. This shift, as well as the one
expressed by the last term in Eq.\ (\ref{GPexpm}), is, however,  of
 the
order of $A^{2}\left( t\right) \epsilon _{0}$, and can only lead to a
 negligibly small shift of the
resonance. Thus, an analysis of an expanding hybrid atom-molecule BEC
 can
be reduced to the solution of a set of ordinary differential equations
(\ref{GPexpa}) and (\ref{GPexpm}). The parametric dependence on
 $\bm{\rho }$
arises from the inhomogeneous Thomas-Fermi initial conditions for
$\Phi _{0}\left( \bm{\rho },t\right) $ expressed by Eq.\
(\ref{ThomasFermi}).

\section{Parametric approximation\label{SecParAppr}}

\subsection{\label{SecDump}Dump state elimination}

Consider an atom-molecule homogeneous quantum gas described by
the second-quantized Hamiltonian (\ref{genH}). Let the initial state
of the atomic field at $t=t_{0}$  be a coherent state of zero kinetic
energy
\begin{equation}
\hat{\Psi }_{a}\left( {\bf p},t_{0}\right) |\text{in}\rangle =\left(
 2\pi \right) ^{3/2}\varphi _{0}\left( t_{0}\right) \delta \left(
 {\bf p}\right) |\text{in}\rangle  , \label{Psiat0}
\end{equation}
where $|\varphi _{0}\left( t_{0}\right) |^{2}=n_{a}\left( t_{0}\right
) $ is the initial atomic density and $|$in$\rangle $ is
the  time-independent state vector in the Heisenberg picture. A pair
of condensate atoms forms a molecule of zero kinetic energy. Therefore
the resonant molecules can be represented by a mean field $\varphi
 _{m}\left( t\right) $ as
\begin{equation}
\langle \text{in}|\hat{\Psi }_{m}\left( {\bf p},t\right)
|\text{in}\rangle =\left( 2\pi \right) ^{3/2}\varphi _{m}\left(
 t\right) \delta \left( {\bf p}\right)  , \label{Psiphim}
\end{equation}
where $|\varphi _{m}\left( t\right) |^{2}=n_{m}\left( t\right) $ is
 the molecular condensate density. This
approach therefore takes into account the time dependence of the
molecular mean field, but neglects fluctuations of the molecular field
due to a Feshbach coupling of non-condensate atoms.

The outcome of atom-molecule and molecule-molecule deactivating
collisions is introduced, as in the mean field theory of the previous
section, by adding the molecular dump states. The elimination of these
states in a second-quantized description should, however, be done in a
different way (see Ref.\ \cite{YB03}). It is similar to the
Heisenberg-Langevin formalism of quantum optics (see Refs.\
\cite{SZ97,BR97}), but takes into account the nonlinearity of the
collisional damping.

The Hamiltonian (\ref{genH}) yields the following equations of
motion for the the annihilation operators $\hat{\Psi }_{a}\left( {\bf
 p},t\right) $ of the atomic field
and $\hat{\Psi }_{d}\left( {\bf p},t\right) $ of the molecular dump
 states,
\begin{eqnarray}
i\dot{\hat{\Psi }}_{a}\left( {\bf p},t\right)  =\left\lbrack {p{ }
 ^{2}\over 2m}+\epsilon _{a}\left( t\right) \right\rbrack \hat{\Psi
 }_{a}\left( {\bf p},t\right)  \nonumber
\\
+2\left( 2\pi \right) ^{-3/2}\int d^{3}p^\prime \tilde{d}_{h}\left(
 {\bf p}-{\bf p}^\prime \right) \hat{\Psi }^{\dag }_{a}\left( {\bf
 p}^\prime ,t\right) \hat{\Psi }_{m}\left( {\bf p}+{\bf p}^\prime
 ,t\right)  \nonumber
\\
+\left( 2\pi \right) ^{-3/2}\sum\limits^{}_{d} \int d^{3}p^\prime
 d^{3}p_{d}\tilde{d}^{*}_{d}\left( {\bf p}_{d}-2{\bf p}^\prime \right
) \hat{\Psi }^{\dag }_{m}\left( {\bf p}^\prime +{\bf p}_{d}-{\bf
 p},t\right) \hat{\Psi }_{a}\left( {\bf p}^\prime ,t\right) \hat{\Psi
 }_{d}\left( {\bf p}_{d},t\right)  \nonumber
\\
+\left( 2\pi \right) ^{-3/2}\sum\limits^{}_{d} \int
 d^{3}p_{d}d^{3}p_{m}\tilde{d}_{d}\left( {\bf p}_{d}-2{\bf p}\right)
 \hat{\Psi }^{\dag }_{d}\left( {\bf p}_{d},t\right) \hat{\Psi
 }_{m}\left( {\bf p}_{m},t\right) \hat{\Psi }_{a}\left( {\bf p}+{\bf
 p}_{d}-{\bf p}_{m},t\right)  \nonumber
\\
\label{Psiad}
\\
i\dot{\hat{\Psi }}_{d}\left( {\bf p}_{d},t\right)  =\left\lbrack {p{
 } ^{2}_{d}\over 4m}-E_{d}\right\rbrack \hat{\Psi }_{d}\left( {\bf
 p}_{d},t\right)  \nonumber
\\
+\left( 2\pi \right) ^{-3/2}\int d^{3}p^\prime
 d^{3}p_{m}\tilde{d}_{d}\left( {\bf p}_{d}-2{\bf p}^\prime \right)
 \hat{\Psi }^{\dag }_{a}\left( {\bf p}^\prime ,t\right) \hat{\Psi
 }_{m}\left( {\bf p}_{m},t\right) \hat{\Psi }_{a}\left( {\bf
 p}^\prime +{\bf p}_{d}-{\bf p}_{m},t\right)  \nonumber
\\
+\left( 2\pi \right) ^{-3/2}\sum\limits^{}_{u} \int
 d^{3}p_{m}d^{3}p_{u}\tilde{d}_{ud}\left( {\bf p}_{u}-{\bf
 p}_{d}\right) \hat{\Psi }^{\dag }_{u}\left( {\bf p}_{u},t\right)
 \hat{\Psi }_{m}\left( {\bf p}_{m},t\right) \hat{\Psi }_{m}\left(
 {\bf p}_{u}+{\bf p}_{d}-{\bf p}_{m},t\right)  \nonumber
\\
\label{Psid}
\end{eqnarray}
The atom and the molecule emerging from the deactivation event
(\ref{AMCol}) depart with momenta $p_{d}\ge  P_{d}$  [see Eq.\
(\ref{Pd})]. The
deactivation energy $E_{d}$  substantially exceeds characteristic
 energies
of atoms formed by dissociation of the condensate molecules, allowing
us to discriminate two groups of atoms, with momenta above and below
$\min\left( P_{d}\right) $, respectively. [This assumption is
 equivalent to the condition
(\ref{ortphi}).] Equations (\ref{Psiad}) and (\ref{Psid}) give the
following equation of motion for the product of the field operators
(with $p_{d}\ge \min\left( P_{d}\right) $)
\begin{eqnarray}
i{\partial \over \partial t}\left\lbrack \hat{\Psi }_{d}\left( {\bf
 p}_{d},t\right) \hat{\Psi }_{a}\left( {\bf p}-{\bf p}_{d},t\right)
 \right\rbrack &&\approx \left\lbrack {p{ } ^{2}_{d}\over 4m}+{\left(
 {\bf p}_{d}-{\bf p}\right) { } ^{2}\over 2m}-E_{d}\right\rbrack
 \hat{\Psi }_{d}\left( {\bf p}_{d},t\right) \hat{\Psi }_{a}\left(
 {\bf p}-{\bf p}_{d},t\right)  \nonumber
\\
&& +\tilde{d}_{d}\left( 3{\bf p}_{d}-2{\bf p}\right) \hat{\Psi
 }_{a}\left( {\bf p},t\right) \varphi _{m}\left( t\right)  ,
 \label{PsidPsia}
\end{eqnarray}
where $\epsilon _{a}$  is neglected as small compared to $E_{d}$, and
 the molecular
field operator $\hat{\Psi }_{m}\left( {\bf p},t\right) $ is replaced
 by the mean field $\varphi _{m}\left( t\right) $ [see Eq.\
(\ref{Psiphim})]. The source term in Eq.\ (\ref{PsidPsia}) arises from
the commutation of field operators upon normal ordering, while the
terms containing the dump field operators are neglected here.
Substitution of the solution of Eq.\ (\ref{PsidPsia}) and the
molecular mean field (\ref{Psiphim}) into Eq.\ (\ref{Psiad}) gives the
following integro-differential equation
\begin{eqnarray}
i\dot{\hat{\Psi }}_{a}\left( {\bf p},t\right)  =\left\lbrack {p{ }
 ^{2}\over 2m}+\epsilon _{a}\left( t\right) \right\rbrack \hat{\Psi
 }_{a}\left( {\bf p},t\right) +2\tilde{d}^{*}_{h}\left( 2p\right)
 \varphi _{m}\left( t\right) \hat{\Psi }^{\dag }_{a}\left( -{\bf
 p},t\right)  \nonumber
\\
-i\varphi ^{*}_{m}\left( t\right)  \int\limits^{t}_{t{ }
 _{0}}dt^\prime  K\left( t-t^\prime \right) \varphi _{m}\left(
 t^\prime \right) \hat{\Psi }_{a}\left( {\bf p},t^\prime \right) +i
 \hat{F}\left( {\bf p},t\right)  \label{QEMNM}
\end{eqnarray}
with a kernel
\begin{equation}
K\left( t-t^\prime \right) =\sum\limits^{}_{d}\int d^{3}p_{d}
|\tilde{d}_{d}\left( 3{\bf p}_{d}-2{\bf p}\right)
|^{2}\exp\left\lbrack -i\left( {p{ } ^{2}_{d}\over 4m}+{\left( {\bf
 p}_{d}-{\bf p}\right) { } ^{2}\over 2m}-E_{d}\right) \left(
 t-t^\prime \right) \right\rbrack  ,
\end{equation}
and a quantum noise source
\begin{eqnarray}
\hat{F}\left( {\bf p},t\right) =-i\varphi ^{*}_{m}\left( t\right)
 \sum\limits^{}_{d}\int d^{3}p_{d}\tilde{d}^{*}_{d}\left( 3{\bf
 p}_{d}-2{\bf p}\right) \hat{\Psi }_{d}\left( {\bf p}_{d},t_{0}\right
) \hat{\Psi }_{a}\left( {\bf p}-{\bf p}_{d},t_{0}\right)  \nonumber
\\
\times \exp\left\lbrack -i\left( {p{ } ^{2}_{d}\over 4m}+{\left( {\bf
 p}_{d}-{\bf p}\right) { } ^{2}\over 2m}-E_{d}\right) \left(
 t-t_{0}\right) \right\rbrack  . \label{Fdef}
\end{eqnarray}
As in the Heisenberg-Langevin formalism, commutators of the
quantum noise are related to the kernel by a fluctuation-dissipation
theorem, except that here this relation involves averages of the
commutators,
\begin{equation}
\langle \text{in}|\left\lbrack \hat{F}\left( {\bf p},t\right)
 ,\hat{F}^{\dag }\left( {\bf p}^\prime ,t^\prime \right)
 \right\rbrack |\text{in}\rangle =\varphi ^{*}_{m}\left( t\right)
 \varphi _{m}\left( t^\prime \right) K\left( t-t^\prime \right)
 \delta \left( {\bf p}-{\bf p}^\prime \right) .
\end{equation}
In the Markovian approximation, the kernel is assumed to be
sharply peaked at $t=t'{}$, so that
\begin{equation}
K\left( t-t'{}\right) ={1\over 2}k_{a}\delta \left( t-t'{}\right)  ,
 \label{MarkAppr}
\end{equation}
where the deactivation rate coefficient $k_{a}$  is defined by Eq.\
(\ref{gamma}). An expression of the kernel by Eq.\ (\ref{MarkAppr})
implies that the system retains no memory of its history. The equation
of motion for the atomic field then attains the Heisenberg-Langevin
form
\begin{equation}
i\dot{\hat{\Psi }}_{a}\left( {\bf p},t\right)  =\left\lbrack {p{ }
 ^{2}\over 2m} + \epsilon _{a}\left( t\right) -i{k{ } _{a}\over 2}
|\varphi _{m}\left( t\right) |^{2}\right\rbrack \hat{\Psi }_{a}\left(
 {\bf p},t\right) +2g^{*}\varphi _{m}\left( t\right) \hat{\Psi
 }^{\dag }_{a}\left( -{\bf p},t\right)  +i \hat{F}\left( {\bf
 p},t\right)  , \label{Psia}
\end{equation}
where $\tilde{d}_{h}\left( p\right) $ is replaced by its maximal
 value $\tilde{d}_{h}\left( 0\right) =g$ [see Eq.\
(\ref{gVd})]. The shift associated with the deactivating collisions
 $\delta _{a}$
[see Eq.\ (\ref{delta})] can be neglected compared to other energy
scales in real physical situations. The quantum noise source
 $\hat{F}\left( {\bf p},t\right) $ is
$\delta $-correlated in the Markovian approximation, obeying
\begin{equation}
\langle \text{in}|\left\lbrack \hat{F}\left( {\bf p},t\right)
 ,\hat{F}^{\dag }\left( {\bf p}^\prime ,t^\prime \right)
 \right\rbrack |\text{in}\rangle =k_{a}|\varphi _{m}\left( t\right)
|^{2}\delta \left( t-t^\prime \right) \delta \left( {\bf p}-{\bf
 p}^\prime \right)  . \label{Fcomm}
\end{equation}
The Markovian approximation is applied only to deactivating
collisions, while the description of the association-dissociation
 processes
remains non-Markovian.

\subsection{Atomic field representation\label{SecAtField}}

Equation (\ref{Psia}) is a linear inhomogeneous operator equation.
Consider at first a solution of the corresponding homogeneous
 equation.
Linear operator equations can be solved as $c$-number ones. A general
solution can be expressed as
\begin{equation}
\hat{\Psi }_{a}\left( {\bf p},t\right) =C\left( t\right) \left\lbrack
 \hat{\Psi }\left( {\bf p},t_{0}\right) \psi _{c}\left( p,t\right)
+\hat{\Psi }^{\dag }\left( -{\bf p},t_{0}\right) \psi _{s}\left(
 p,t\right) \right\rbrack , \label{PsiaHom}
\end{equation}
in terms of the creation and annihilation operators at the initial
time $t=t_{0}$  and the solutions $\psi _{c,s}\left( p,t\right) $ of
 the corresponding  $c$-number
equations
\begin{equation}
i\dot{\psi }_{c,s}\left( p,t\right) =\left\lbrack {p{ } ^{2}\over 2m}
 + \epsilon _{a}\left( t\right) \right\rbrack  \psi _{c,s}\left(
 p,t\right) +2g^{*}\varphi _{m}\left( t\right) \psi ^{*}_{s,c}\left(
 p,t\right)  , \label{Psics}
\end{equation}
given the initial conditions $\psi _{c}\left( p,t_{0}\right) =1$,
 $\psi _{s}\left( p,t_{0}\right) =0$. Equation
(\ref{Psics}) as well as the functions $\psi _{c,s}\left( p,t\right)
 $ are independent of
the ${\bf p}$ direction. The factor $C\left( t\right) $ takes into
 account the imaginary term
in Eq.\ (\ref{Psia}),
\begin{equation}
C\left( t\right) =\exp\left( -\int\limits^{t}_{t{ } _{0}}d t_{1}{k{ }
 _{a}\over 2}|\varphi _{m}\left( t_{1}\right) |^{2}\right)  .
\end{equation}
A general method for solving inhomogeneous equations is the
variation of constants in the solution of the corresponding
homogeneous equations. Following this method, let us replace the field
operators at $t=t_{0}$, which play the role of constants in the
 solution
(\ref{PsiaHom}), by unknown time-dependent operators $\hat{A}\left(
 {\bf p},t\right) $,
representing the atomic field operator in the form
\begin{equation}
\hat{\Psi }_{a}\left( {\bf p},t\right) =C\left( t\right) \left\lbrack
 \hat{A}\left( {\bf p},t\right) \psi _{c}\left( p,t\right)
+\hat{A}^{\dag }\left( -{\bf p},t\right) \psi _{s}\left( p,t\right)
 \right\rbrack . \label{PsiaA}
\end{equation}
The initial conditions at $t=t_{0}$  require that
\begin{equation}
\hat{A}\left( {\bf p},t_{0}\right) =\hat{\Psi }_{a}\left( {\bf
 p},t_{0}\right)  .
\end{equation}
Substitution of Eq.\ (\ref{PsiaA}) into Eq.\ (\ref{Psia}) leads to
the equation
\begin{equation}
\dot{\hat{A}}\left( {\bf p},t\right) \psi _{c}\left( p,t\right)
+\dot{\hat{A}}^{\dag }\left( -{\bf p},t\right) \psi _{s}\left(
 p,t\right) =\hat{F}\left( {\bf p},t\right) C^{-1}\left( t\right)  ,
\end{equation}
which, together with its hermitian conjugate, form a system of
linear equations for $\dot{\hat{A}}\left( {\bf p},t\right) $ and
 $\dot{\hat{A}}^{\dag }\left( -{\bf p},t\right) $ with the determinate
\begin{equation}
|\psi _{c}\left( p,t\right) |^{2}-|\psi _{s}\left( p,t\right) |^{2}=1
 ,
\end{equation}
as one can prove by using of Eq.\ (\ref{Psics}). As a result, the
operator $\hat{A}\left( {\bf p},t\right) $ can be expressed in the
 form
\begin{equation}
\hat{A}\left( {\bf p},t\right) =\hat{\Psi }_{a}\left( {\bf
 p},t_{0}\right) +\int\limits^{t}_{t{ } _{0}}{dt{ } _{1}\over C\left(
 t_{1}\right) }\bigl\lbrack \psi ^{*}_{c}\left( p,t_{1}\right)
 \hat{F}\left( {\bf p},t_{1}\right) -\psi _{s}\left( p,t_{1}\right)
 \hat{F}^{\dag }\left( -{\bf p},t_{1}\right) \bigr\rbrack  . \label{A}
\end{equation}
The reactions (\ref{AMCol}) of deactivating collisions thus
contribute to the description of the atomic field operator both the
factor $C\left( t\right) $ in Eq.\ (\ref{PsiaA}) and the second term,
 containing the
quantum noise, in Eq.\ (\ref{A}). The factor $C\left( t\right) $
 describes the decay
due to deactivating collisions, while the quantum noise provides for
maintaining the correct commutation relations of the atomic field
operators (in an average sense) as
\begin{equation}
\langle \text{in}|\left\lbrack \hat{\Psi }_{a}\left( {\bf p},t\right)
 ,\hat{\Psi }^{\dag }_{a}\left( {\bf p}^\prime ,t\right)
 \right\rbrack |\text{in}\rangle =\delta \left( {\bf p}-{\bf
 p}^\prime \right)  .
\end{equation}
\subsection{The atomic mean field and correlation
 functions\label{SecCorrFun}}

Physical observables are expressed in terms of averages of the field
operators and their products. The atomic field operator includes a
contribution proportional to the quantum noise $\hat{F}\left( {\bf
 p},t\right) $ [see Eqs.\
(\ref{Psia}) and (\ref{A})]. Since the initial state (described by the
state vector $|$in$\rangle $) does not contain the molecules
 A$_{2}\left( d\right) $ and atoms with
momenta $p>P_{d}$, the definition (\ref{Fdef}) of $\hat{F}\left( {\bf
 p},t\right) $ leads to the following
zero-valued averages
\begin{equation}
\langle \text{in}|\hat{F}\left( {\bf p},t\right) |\text{in}\rangle
 =\langle \text{in}|\hat{F}\left( {\bf p},t\right) \hat{F}\left( {\bf
 p}'{},t'{}\right) |\text{in}\rangle =\langle \text{in}|\hat{F}^{\dag
 }\left( {\bf p},t\right) \hat{F}\left( {\bf p}'{},t'{}\right)
|\text{in}\rangle =0 . \label{Faver}
\end{equation}
Averages of products of the quantum noise and atomic field operators
are zero-valued as well.

The commutation relation (\ref{Fcomm}) thus gives
\begin{equation}
\langle \text{in}|\hat{F}\left( {\bf p},t\right) \hat{F}^{\dag }\left
( {\bf p}^\prime ,t^\prime \right) |\text{in}\rangle =k_{a}|\varphi
 _{m}\left( t\right) |^{2}\delta \left( t-t^\prime \right) \delta
 \left( {\bf p}-{\bf p}^\prime \right)  .
\end{equation}
The average of the atomic field operator is determined by Eqs.\
(\ref{Psiat0}), (\ref{A}), and (\ref{PsiaA}) as
\begin{equation}
\langle \text{in}|\hat{\Psi }_{a}\left( {\bf p},t\right)
|\text{in}\rangle =\left( 2\pi \right) ^{3/2}\varphi _{0}\left(
 t\right) \delta \left( {\bf p}\right)  ,
\end{equation}
where
\begin{equation}
\varphi _{0}\left( t\right) =C\left( t\right) \left\lbrack \psi
 _{c}\left( 0,t\right) \varphi _{0}\left( t_{0}\right) +\psi
 _{s}\left( 0,t\right) \varphi ^{*}_{0}\left( t_{0}\right)
 \right\rbrack  \label{phi0p}
\end{equation}
is the atomic condensate mean field.

The averages of the quantum noise lead to the following expressions
for the two-atom correlation functions
\begin{eqnarray}
\langle \text{in}|\hat{\Psi }^{\dag }_{a}\left( {\bf p},t\right)
 \hat{\Psi }_{a}\left( {\bf p}^\prime ,t\right) |\text{in}\rangle
 =\left( 2\pi \right) ^{3}n_{0}\left( t\right) \delta \left( {\bf
 p}\right) \delta \left( {\bf p}^\prime \right) +n_{s}\left(
 p,t\right) \delta \left( {\bf p}-{\bf p}^\prime \right)
 \label{ncorr}
\\
\langle \text{in}|\hat{\Psi }_{a}\left( {\bf p},t\right) \hat{\Psi
 }_{a}\left( {\bf p}^\prime ,t\right) |\text{in}\rangle =\left( 2\pi
 \right) ^{3}m_{0}\left( t\right) \delta \left( {\bf p}\right) \delta
 \left( {\bf p}^\prime \right) +m_{s}\left( p,t\right) \delta \left(
 {\bf p}+{\bf p}^\prime \right)  \label{acorr} ,
\end{eqnarray}
where
\begin{equation}
n_{0}\left( t\right) =|\varphi _{0}\left( t\right) |^{2} \label{n0}
\end{equation}
is the condensate density,
\begin{equation}
n_{s}\left( p,t\right) =|\psi _{s}\left( p,t\right) |^{2}\left\lbrack
 1+\eta _{s}\left( p,t\right) \right\rbrack +|\psi _{c}\left(
 p,t\right) |^{2}\eta _{s}\left( p,t\right) -2\text{Re}\left\lbrack
 \psi ^{*}_{s}\left( p,t\right) \psi _{c}\left( p,t\right) \eta
 _{c}\left( p,t\right) \right\rbrack  \label{ns}
\end{equation}
is the momentum spectrum of the non-condensate atoms, and
\begin{eqnarray}
m_{0}\left( t\right) &=&\varphi ^{2}_{0}\left( t\right)  \nonumber
\\
\label{Anden}
\\
m_{s}\left( p,t\right) &=&\psi _{s}\left( p,t\right) \psi _{c}\left(
 p,t\right) \left\lbrack 1+2\eta _{s}\left( p,t\right) \right\rbrack
 -\psi ^{2}_{c}\left( p,t\right) \eta _{c}\left( p,t\right) -\psi
 ^{2}_{s}\left( p,t\right) \eta ^{*}_{c}\left( p,t\right)  \nonumber
\end{eqnarray}
are the anomalous densities of the condensate and non-condensate
atoms. The functions
\begin{eqnarray}
\eta _{s}\left( p,t\right) =k_{a}C^{2}\left( t\right)
 \int\limits^{t}_{t{ } _{0}}{d t^\prime \over C^{2}\left( t^\prime
 \right) } |\varphi _{m}\left( t^\prime \right) \psi _{s}\left(
 p,t^\prime \right) |^{2} \nonumber
\\
{}\label{etacs}
\\
\eta _{c}\left( p,t\right) =k_{a}C^{2}\left( t\right)
 \int\limits^{t}_{t{ } _{0}}{d t^\prime \over C^{2}\left( t^\prime
 \right) }|\varphi _{m}\left( t^\prime \right) |^{2}\psi _{s}\left(
 p,t^\prime \right) \psi ^{*}_{c}\left( p,t^\prime \right)  \nonumber
\end{eqnarray}
describe the contribution of quantum noise.

The atomic density
\begin{equation}
n_{a}\left( t\right) =\left( 2\pi \right) ^{-3}\int
 d^{3}p_{1}d^{3}p_{2}\exp\left\lbrack i\left( {\bf p}_{2}-{\bf
 p}_{1}\right) {\bf r}\right\rbrack \langle \text{in}|\hat{\Psi
 }^{\dag }_{a}\left( {\bf p}_{1},t\right) \hat{\Psi }_{a}\left( {\bf
 p}_{2},t\right) |\text{in}\rangle
\end{equation}
then appears to be ${\bf r}$-independent, and comprises the sum
\begin{equation}
n_{a}\left( t\right) =n_{0}\left( t\right) +n_{s}\left( t\right)
 \label{adens}
\end{equation}
of the densities of condensate atoms $n_{0}\left( t\right) $  [see
 Eq.\
(\ref{n0})],  and of non-condensate (entangled) atoms $n_{s}\left(
 t\right) $
in a wide spectrum of kinetic energies $E=p^{2}/\left( 2m\right) $,
\begin{equation}
n_{s}\left( t\right) =\int dE \tilde{n}_{s}\left( E,t\right)  ,
 \label{nsE}
\end{equation}
where the energy spectrum $\tilde{n}_{s}\left( E,t\right) $ is
 related to the momentum
spectrum $n_{s}\left( p,t\right) $ [see Eq.\ (\ref{ns})] as
\begin{equation}
\tilde{n}_{s}\left( E,t\right) ={m p\over 2\pi { } ^{2}}n_{s}\left(
 p,t\right)  . \label{EnergySpect}
\end{equation}
\subsection{The molecular field\label{SecMolField}}

The equation of motion for the molecular field operator $\hat{\Psi
 }_{m}\left( {\bf p},t\right) $
is obtained by a procedure of dump field elimination which is similar
to the one presented in Sec.\ \ref{SecDump} for the atomic field.
After substitution of the solution of Eq.\ (\ref{PsidPsia}), and a
similar equation for the product $\hat{\Psi }_{d}\left( {\bf
 p}_{d},t\right) \hat{\Psi }_{u}\left( {\bf p}_{u},t\right) $, the
 operator
equation of motion attains the form:
\begin{eqnarray}
i\dot{\hat{\Psi }}_{m}\left( {\bf p}_{m},t\right) ={p{ }
 ^{2}_{m}\over 4m}\hat{\Psi }_{m}\left( {\bf p}_{m},t\right) +\left(
 2\pi \right) ^{-3/2}\int d^{3}p \tilde{d}_{h}\left( 2{\bf p}-{\bf
 p}_{m}\right) \hat{\Psi }_{a}\left( {\bf p},t\right) \hat{\Psi
 }_{a}\left( {\bf p}_{m}-{\bf p},t\right)  \nonumber
\\
-i\left( 2\pi \right) ^{-3/2}\int\limits^{t}_{t{ } _{0}}dt^\prime
 \int d^{3}p K\left( t-t^\prime \right) \hat{\Psi }^{\dag }_{a}\left(
 {\bf p}-{\bf p}_{m},t\right) \hat{\Psi }_{a}\left( {\bf p},t^\prime
 \right) \varphi _{m}\left( t'{}\right)  \nonumber
\\
-i\left( 2\pi \right) ^{3/2}\int\limits^{t}_{t{ } _{0}}dt^\prime
 K_{m}\left( t-t^\prime \right) \varphi ^{*}_{m}\left( t\right)
 \varphi ^{2}_{m}\left( t'{}\right) \delta \left( {\bf p}_{m}\right)
+i \hat{F}_{a}\left( {\bf p},t\right) +i \hat{F}_{m}\left( {\bf
 p},t\right)  ,
\end{eqnarray}
where the kernel $K_{m}\left( t-t^\prime \right) $ pertains to
 molecule-molecule
deactivating collisions (\ref{MMCol}). In the Markovian approximation
it is approximated by $K_{m}\left( t-t'{}\right) =k_{m}\delta \left(
 t-t'{}\right) $ where the deactivation rate
$k_{m}$  is defined by Eq.\ (\ref{gammam}). The quantum noise sources
$\hat{F}_{a}\left( {\bf p},t\right) $ and $\hat{F}_{m}\left( {\bf
 p},t\right) $, related to atom-molecule and molecule-molecule
deactivation, vanish on mean-field averaging like $\hat{F}\left( {\bf
 p},t\right) $ [see Eq.\
(\ref{Faver})]. The right-hand side of the resulting equation is
proportional to $\delta \left( {\bf p}_{m}\right) $ by virtue of Eqs.
 \ (\ref{ncorr}) and
(\ref{acorr}), thus securing the consistency of Eq.\ \
(\ref{Psiphim}). Taking into account Eq.\ (\ref{MarkAppr}) we obtain
the equation of motion for the molecular mean field
\begin{equation}
i \dot{\varphi }_{m}\left( t\right) =g m_{0}\left( t\right) +{1\over
 2\pi { } ^{2}}\int\limits^{\infty }_{0}p^{2}d p \tilde{d}_{h}\left(
 2{\bf p}\right) m_{s}\left( p,t\right) -i\left( {k{ } _{a}\over
 2}n_{a}\left( t\right) +k_{m}|\varphi _{m}\left( t\right)
|^{2}\right) \varphi _{m}\left( t\right) , \label{Phim}
\end{equation}
where the the anomalous densities of the condensate and
non-condensate atoms, $m_{0}\left( t\right) $ and $m_{s}\left(
 p,t\right) $, are defined by Eq.\
(\ref{Anden}) and the atomic density $n_{a}$  is defined by Eq.\
(\ref{adens}).  These equations can describe the spontaneous
dissociation of molecules into the atomic vacuum ($m_{0}=0$) due to
the term in Eq.\ (\ref{Phim}) proportional to $m_{s}$.

Equation (\ref{Phim}) cannot be solved with a constant
 $\tilde{d}_{h}\left( {\bf p}\right) $ due
to the divergence of the integral over $p$. It can be seen from
asymptotic behavior of the anomalous density $m_{s}\left( p,t\right)
 $ as $p\rightarrow \infty $. In this
limit the solutions of Eq.\ (\ref{Psics}) can be approximated as
\begin{eqnarray}
\psi _{c}\left( p,t\right) \sim \exp\left( -i{p{ } ^{2}\over 2m}\left
( t-t_{0}\right) \right)  \nonumber
\\
\label{psicsas}
\\
\psi _{s}\left( p,t\right) \sim -{2mg{ } ^{*}\over p{ }
 ^{2}}\left\lbrack \varphi _{m}\left( t\right) \exp\left( i{p{ }
 ^{2}\over 2m}\left( t-t_{0}\right) \right) -\varphi _{m}\left(
 t_{0}\right) \exp\left( -i{p{ } ^{2}\over 2m}\left( t-t_{0}\right)
 \right) \right\rbrack  , \nonumber
\end{eqnarray}
leading to
\[
m_{s}\left( p,t\right) \sim -{2mg{ } ^{*}\over p{ } ^{2}}\varphi
 _{m}\left( t\right)  \nonumber
\]
and hence to an asymptotically non-decreasing integrand in Eq.\
(\ref{Phim}). Following Ref.\ \cite{HPW01} let us introduce a momentum
cut-off $p_{c}$,  setting $\tilde{d}_{h}\left( 2{\bf p}\right) =g$
 for $p<p_{c}$  and $\tilde{d}_{h}\left( 2{\bf p}\right) =0$
 otherwise, where $g$ is
defined by Eq.\ (\ref{Delta}). The integral term in Eq.\ (\ref{Phim})
tends to $E_{m}\varphi _{m}$  in the limit $p_{c}\rightarrow \infty $
 with $E_{m}=-{m\over \pi { } ^{2}}|g|^{2}p_{c}$. Substituting
renormalized functions
\begin{eqnarray}
\varphi _{m}\left( t\right) \leftarrow\varphi _{m}\left( t\right)
 \exp\left( -i E_{m}t\right) ,\qquad \psi _{c,s}\left( p,t\right)
 \leftarrow\psi _{c,s}\left( p,t\right) \exp\left( -{i\over 2}
 E_{m}t\right)  \nonumber
\\
\epsilon _{a}\left( t\right) \leftarrow\epsilon _{a}\left( t\right)
+{1\over 2}E_{m} \nonumber
\end{eqnarray}
into Eqs.\ (\ref{Psics}) and (\ref{Phim}) one obtains a
non-divergent equation for the molecular mean field
\begin{eqnarray}
i \dot{\varphi }_{m}\left( t\right) =g m_{0}\left( t\right) +{1\over
 2\pi { } ^{2}}\int\limits^{p{ } _{c}}_{0}d p \left\lbrack p^{2}g
 m_{s}\left( p,t\right) +2m|g|^{2}\varphi _{m}\left( t\right)
 \right\rbrack  \nonumber
\\
-i\left( {k{ } _{a}\over 2}n_{a}\left( t\right) +k_{m}|\varphi
 _{m}\left( t\right) |^{2}\right) \varphi _{m}\left( t\right) ,
 \label{Phimr}
\end{eqnarray}
while Eq.\ (\ref{Psics}) retains its form. A numerical solution
of Eqs.\ (\ref{Psics}) on a grid of values of $p$, combined with Eq.\
(\ref{Phimr}), is consistently sufficient for elucidating the dynamics
of the system.

Numerical calculations and a qualitative analysis can be made
more convenient by using dimensionless variables. Let the time and
momentum be rescaled to
\begin{equation}
\tilde{t}=|g|\sqrt{n}t,\qquad \tilde{p}=\left( m|g|\right) ^{-1
/2}n^{-1/4}p,
\end{equation}
where the density scale
\begin{equation}
n=n_{a}\left( t_{0}\right) +2n_{m}\left( t_{0}\right)
\end{equation}
is determined in terms of the initial atomic and molecular
densities. The scale for all energies is thus $|g|\sqrt{n}$. The
 deactivation
rate coefficients are rescaled to
\begin{equation}
\tilde{k}_{a,m}={\sqrt{n}\over |g|}k_{a,m} . \label{kamresc}
\end{equation}
The equation of motion for the rescaled molecular mean field
attains the form
\begin{eqnarray}
i \dot{\tilde{\varphi }}_{m}\left( \tilde{t}\right)
 =\tilde{m}_{0}\left( \tilde{t}\right) +\sigma
 \int\limits^{\tilde{p}{ } _{c}}_{0}d \tilde{p} \left\lbrack
 \tilde{p}^{2}\tilde{m}_{s}\left( \tilde{p},t\right) +2\tilde{\varphi
 }_{m}\left( \tilde{t}\right) \right\rbrack  \nonumber
\\
-i\left\lbrack {\tilde{k}{ } _{a}\over 2}\tilde{n}_{0}\left(
 \tilde{t}\right) +{\tilde{k}{ } _{a}\over 2}\sigma
 \int\limits^{\tilde{p}{ } _{c}}_{0}d \tilde{p}
 \tilde{p}^{2}\tilde{n}_{s}\left( \tilde{p},t\right) +\tilde{k}_{m}
|\tilde{\varphi }_{m}\left( \tilde{t}\right) |^{2}\right\rbrack
 \tilde{\varphi }_{m}\left( \tilde{t}\right) , \label{Phimrd}
\end{eqnarray}
where the rescaled densities $\tilde{n}_{0}$, $\tilde{n}_{s}$,
 $\tilde{m}_{0}$, and $\tilde{m}_{s}$  are determined by the
same equations as the non-rescaled ones, but expressed in terms of
 rescaled
parameters. The dimensionless coefficient $\sigma $ is given by
\begin{equation}
\sigma ={\left( m|g|\right) { } ^{3/2}\over 2\pi ^{2}n{ } ^{1
/4}}=\left\lbrack {\left( |a_{a}|m\mu \Delta \right) { } ^{3}\over
 2\pi ^{5}n}\right\rbrack ^{1/4} , \label{sigma}
\end{equation}
expressed in terms of the resonance strength and initial densities.
 The
contribution of non-condensate atoms (atomic field fluctuations) to
 Eq.\
(\ref{Phimrd}) is proportional to $\sigma $ and, therefore, increases
 with the
resonance strength and is suppressed by high densities. In contrast,
 the
role of fluctuations in a pure atomic gas increases with the density,
according to the Bogoliubov theory (see Ref.\ \cite{DGPS99}).

\subsection{Relation to the Hartree-Fock-Bogoliubov
approach\label{SecHFB}}

The parametric approximation takes into account second order
correlations of the atomic field. The Hartree-Fock-Bogoliubov
approach of Ref.\ \cite{HPW01} takes into account the atomic field
correlations to the same order, but neglects the relaxation due to
deactivating collisions. It consists of a solution of the equations
of motion for the atomic and molecular condensate mean fields
 $\varphi _{a}\left( t\right) $
and $\varphi _{m}\left( t\right) $, as well as the normal and
 anomalous densities of the
non-condensate atoms, $n_{s}\left( p,t\right) $  and $m_{s}\left(
 p,t\right) $. A conventional
derivation of the Hartree-Fock-Bogoliubov equations is based on the
Heisenberg equation of motion for the atomic field operator. Use of
the Heisenberg-Langevin equation (\ref{Psia}) allows to include the
relaxation process. The definitions of $\varphi _{a}\left( t\right)
 $, $n_{s}\left( p,t\right) $,  and $m_{s}\left( p,t\right) $
[Eqs.\ (\ref{phi0p}), (\ref{ns}), and (\ref{Anden}), respectively]
in a combination with Eqs.\ (\ref{Psics}) and (\ref{etacs}) lead,
after some algebra, to the following equations of motion,
\begin{eqnarray}
&&i \dot{\varphi }_{0}\left( t\right) =\left\lbrack \epsilon
 _{a}\left( t\right) -{i\over 2}k_{a}|\varphi _{m}\left( t\right)
|^{2}\right\rbrack \varphi _{0}\left( t\right) +2g^{*}\varphi ^{
*}_{0}\left( t\right) \varphi _{m}\left( t\right)  \nonumber
\\
&&\dot{n}_{s}\left( p,t\right) =-4\text{Im}\left\lbrack g\varphi ^{
*}_{m}\left( t\right) m_{s}\left( p,t\right) \right\rbrack -k_{a}
|\varphi _{m}\left( t\right) |^{2}n_{s}\left( p,t\right)  \label{HFB}
\\
&&i \dot{m}_{s}\left( p,t\right) =\left\lbrack {p{ } ^{2}\over m} +
 2\epsilon _{a}\left( t\right) -i k_{a}|\varphi _{m}\left( t\right)
|^{2}\right\rbrack m_{s}\left( p,t\right) +2g^{*}\varphi _{m}\left(
 t\right) \left\lbrack 1+2n_{s}\left( p,t\right) \right\rbrack .
 \nonumber
\end{eqnarray}
When the deactivating collisions are  neglected ($k_{a}=k_{m}=0$),
 these
equations, in combination with (\ref{Phim}) are just the momentum
representation of the Hartree-Fock-Bogoliubov equations used in
\cite{HPW01} and nothing else, leaving the present approach
mathematically equivalent to the Hartree-Fock-Bogoliubov one. In the
absence of relaxation we have $\eta _{c,s}\left( p,t\right) =0$ and
 the normal and anomalous
densities are expressed as
\begin{equation}
n_{s}\left( p,t\right) =|\psi _{s}\left( p,t\right) |^{2} ,\qquad
 m_{s}\left( p,t\right) =\psi _{s}\left( p,t\right) \psi _{c}\left(
 p,t\right)  . \label{nsms}
\end{equation}
Therefore a solution of equations for $\psi _{c,s}\left( p,t\right) $
 has some benefits
for numerical calculations due to less variation of these functions
compared to $n_{s}\left( p,t\right) $ and $m_{s}\left( p,t\right) $.
 Moreover, Eq.\ (\ref{Psics}) is much
easier to analyze, and can even produce analytical solutions in some
situations.

\subsection{Exactly soluble models\label{SecExSol}}

Consider the dissociation of a molecular  BEC neglecting
deactivating collisions ($k_{a}=k_{m}=0$) and the depletion of the
 molecular
field in comparison to its initial value, so that one can
approximately assume $\varphi _{m}=$const. In this case the various
 atomic ${\bf p}$-modes
become decoupled and Eq.\ (\ref{Psia}) can be solved separately for
each mode. The resulting equations
\begin{equation}
i\dot{\hat{\Psi }}_{a}\left( {\bf p},t\right)  = \epsilon _{p}\left(
 t\right) \hat{\Psi }_{a}\left( {\bf p},t\right) +2\tilde{g}^{
*}\hat{\Psi }^{\dag }_{a}\left( -{\bf p},t\right)  , \label{PsiaM}
\end{equation}
where $\epsilon _{p}\left( t\right) = {p{ } ^{2}\over 2m} + \epsilon
 _{a}\left( t\right) $ and $\tilde{g}=g\varphi ^{*}_{m}$, are similar
 to the equations for
non-adiabatic transitions in two-state systems (see Refs.\
\cite{ChildMCT,NU84,DO88,Akulin05}). However, here we have a
 transition
between the atom creation operators $\hat{\Psi }^{\dag }_{a}\left(
 -{\bf p},t\right) $ and the annihilation (or
hole-creation) one  $\hat{\Psi }_{a}\left( {\bf p},t\right) $. Thus,
 a transition from a hole to an atom
corresponds to the formation of an atomic pair as a result of
 molecular
dissociation. Since the field operators are non-hermitian, the
 many-body
non-adiabatic transition problem (in contrast to a single-body problem
involving a single molecule) has an antihermitian coupling matrix,
 leading
to complex adiabatic energies.

The atomic field operator can be expressed in the form
(\ref{PsiaHom}) with $C=1$. The equations of motion for the
 coefficients
$\psi _{c,s}\left( p,t\right) $ can be represented as
\begin{equation}
i\dot{\psi }_{c,s}\left( p,t\right) =\epsilon _{p}\left( t\right)
 \psi _{c,s}\left( p,t\right) +2\tilde{g}^{*}\psi ^{*}_{s,c}\left(
 p,t\right)  , \label{PsicsM}
\end{equation}
leaving the initial conditions unchanged,
\begin{equation}
\psi _{c}\left( p,t_{0}\right) =1,\qquad \psi _{s}\left(
 p,t_{0}\right) =0. \label{psicpsist0}
\end{equation}
The first-order coupled equations (\ref{PsicsM}) for $\psi _{c}\left(
 p,t\right) $  and
$\varphi _{s}\left( p,t\right) $  can be transformed to second-order
 uncoupled equations
\begin{equation}
\ddot{\psi }_{c,s}\left( p,t\right)  +\left( \epsilon ^{2}_{p}\left(
 t\right) -4|\tilde{g}|^{2}+i\dot{\epsilon }_{p}\left( t\right)
 \right)  \psi _{c,s}\left( p,t\right) =0 , \label{pside2}
\end{equation}
with the same initial conditions (\ref{psicpsist0}) for the two
solutions, augmented by initial conditions  for the derivatives,
\begin{equation}
\dot{\psi }_{c}\left( t_{0}\right) =-i\epsilon _{p}\left( t_{0}\right
) ,\qquad \dot{\psi }_{s}\left( t_{0}\right) =-2i\tilde{g}^{*} ,
 \label{dpsit0}
\end{equation}
derivable from Eqs.\ (\ref{PsicsM}) at $t=t_{0}$. The solutions
$\psi _{c}\left( p,t\right) $  and $\varphi _{s}\left( p,t\right) $
 can be expressed in terms of a pair of
conventionally selected independent solutions of Eq.\ (\ref{pside2})
$\psi _{1}\left( t\right) $ and $\psi _{2}\left( t\right) $, treated
 as ``standard solutions'', in the form

\leftline{$\psi _{c}\left( t\right) ={i\over W\{\psi _{1},\psi
 _{2}\}}\left\lbrack \left( \epsilon \left( t_{0}\right) \psi
 _{2}\left( t_{0}\right) -i\dot{\psi }_{2}\left( t_{0}\right) \right)
 \psi _{1}\left( t\right) -\left( \epsilon \left( t_{0}\right) \psi
 _{1}\left( t_{0}\right) -i\dot{\psi }_{1}\left( t_{0}\right) \right)
 \psi _{2}\left( t\right) \right\rbrack $}
\begin{equation}
\psi _{s}\left( t\right) ={2ig{ } ^{*}\over W\{\psi _{1},\psi
 _{2}\}}\left\lbrack \psi _{2}\left( t_{0}\right) \psi _{1}\left(
 t\right) -\psi _{1}\left( t_{0}\right) \psi _{2}\left( t\right)
 \right\rbrack  , \label{psis}
\end{equation}
where
\begin{equation}
W\{\psi _{1},\psi _{2}\}=\psi _{1}\dot{\psi }_{2}-\psi _{2}\dot{\psi
 }_{1}
\end{equation}
is the Wronskian of the standard solutions.

Consider at first a case of a time-independent energy $\epsilon
 _{p}=$const.
The standard solutions are $\psi _{1,2}\left( t\right) =\exp\left(
 \pm \kappa t\right) $ with
\begin{equation}
\kappa =\sqrt{4|\tilde{g}|^{2}-\epsilon { } ^{2}_{p}} , \label{kappa}
\end{equation}
and $\psi _{c,s}$  attain the form (see Ref.\ \cite{VYA01})
\begin{equation}
\psi _{c}\left( p,t\right) =\cosh \kappa \left( t-t_{0}\right)
 -i{\epsilon { } _{p}\over \kappa }\sinh \kappa \left( t-t_{0}\right)
 ,\qquad \psi _{s}\left( p,t\right) =-i{2\tilde{g}\over \kappa }\sinh
 \kappa \left( t-t_{0}\right)  . \label{epsconst}
\end{equation}
The modes with $\epsilon _{p}<2|\tilde{g}|$, or real $\kappa $, are
 unstable and grow exponentially
at sufficiently long times. This growth is related to Bose enhanced
dissociation and is similar to the amplification of spontaneous
 radiation
in laser-active media, where the parameter $\kappa $ plays the role
 of the
amplification coefficient. A similar solution has been obtained for a
parametric oscillator in quantum optics, see Ref.\ \cite{BR97}.

If a time-dependent $\epsilon _{p}\left( t\right) $ has a real zero,
 the energies of the hole
and atom cross and we face a many-body curve-crossing problem (an
 analog
of a single-body curve crossing involving a single molecule, see
 Refs.\
\cite{ChildMCT,NU84,DO88,Akulin05}). The simplest curve crossing
 model is
the linear one with a time dependence $\epsilon _{p}\left( t\right)
 =-\beta t$, $\beta >0$ (see Ref.\
\cite{YBJ02}). The second-order equation (\ref{pside2}) attains in
 this
case the form of a parabolic cylinder equation (see Ref.\
\cite{Abramovitz})
\begin{equation}
\ddot{\psi }_{c,s}\left( p,t\right)  +\left( \beta ^{2}t^{2}-4
|\tilde{g}|^{2}-i\beta \right)  \psi _{c,s}\left( p,t\right) =0 ,
 \label{pside2l}
\end{equation}
and the standard solutions are expressed in terms of the parabolic
cylinder function (see Ref.\ \cite{YBJ02}). In the limit
 $t_{0}\rightarrow -\infty $, $t\rightarrow \infty $, the
number of non-condensate atoms formed by the spontaneous process can
 be
written as
\begin{equation}
n_{s}\left( p,t\right) \approx \left( 1-e^{-2\pi \lambda }\right)
 e^{2\pi \lambda } , \label{spd0}
\end{equation}
where

$\lambda =2|\tilde{g}|^{2}/\beta  . \label{lambda}$

The parameter  $\lambda $ is nothing else but the Landau-Zener
 exponent (see
Refs.\ \cite{ChildMCT,NU84,DO88,Akulin05}), since the coupling of
 states,
according to Eq.\ (\ref{PsiaM}), is $2\tilde{g}$ and the slope of the
 two-atom energy
is $2\beta $.

The first factor in Eq.\ (\ref{spd0}) is the familiar Landau-Zener
probability, while the second one describes an amplification of the
spontaneous dissociation due to Bose-enhancement. As in Eq.\
(\ref{epsconst}) this amplification is reminiscent of lasing, and the
exponent can be interpreted as a product of a characteristic crossing
 time
($\tilde{g}/\beta $) and an ``amplification coefficient'' $\tilde{g}$.

\begin{figure}
\epsfxsize=0.8\textwidth  \epsfbox{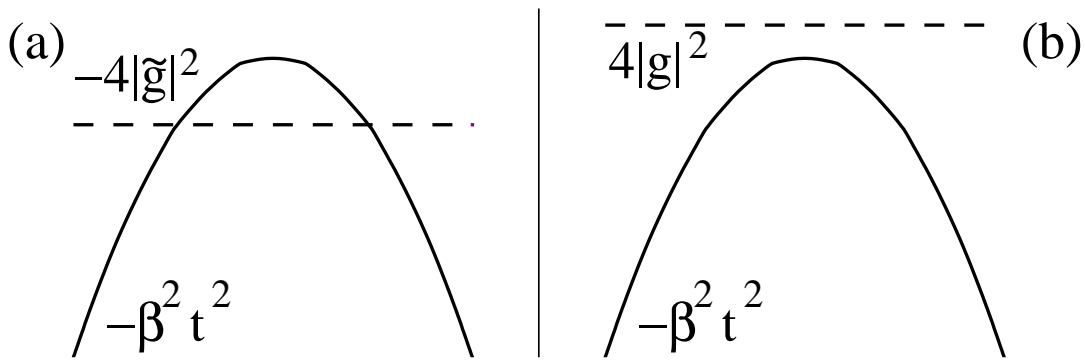}

\caption{Scattering on a parabolic potential barrier
corresponding to the many-body crossing (a, barrier
penetration) vs. single-body crossing (b, above-barrier
reflection). The dashed line marks the ``collision'' energy.
\label{FigBar}}

\end{figure}

A distinction between Eq.\ (\ref{spd0}) and the
Landau-Zener probability can be clarified by interpretation
of the crossing as scattering on a parabolic potential
barrier. Indeed Eq.\ (\ref{pside2l}) can be represented in
the form of one-dimensional stationary Schr\"odinger equation
\begin{equation}
-\ddot{\psi }_{c,s}\left( p,t\right)  -\left( \beta ^{2}t^{2}-i\beta
 \right)  \psi _{c,s}\left( p,t\right) =-4|\tilde{g}|^{2}\psi
 _{c,s}\left( p,t\right)  , \label{ParBar}
\end{equation}
where the time plays a role of a coordinate and the ``energy'$'{}
-4|\tilde{g}|^{2}$  is negative and lies below the the barrier (see
 Fig.\
\ref{FigBar}a). The sign of the energy is a consequence of
antihermiticity of the coupling matrix [see discussion following to
Eq.\ (\ref{PsiaM})]. A single-body curve crossing problem has a
hermitian coupling matrix. It can be described by the same
 Schr\"odinger
equation (\ref{ParBar}) but the energy has an opposite sign and lies
above the barrier (see Fig.\ \ref{FigBar}b). Therefore the many-body
problem corresponds to barrier penetration with exponentially small
penetration probability and large reflection in contrast to large
penetration and exponentially small reflection corresponding to
single-body crossing. This leads to the replacement of the negative
exponents by positive ones, and yields Eq.\ (\ref{spd0}) in place of
the Landau-Zener formula.

The method of Ref.\ \cite{YBJ02} has been applied to a case of
heteronuclear molecules in Ref.\ \cite{KS03}. Exact solutions for some
other shapes of $\epsilon _{a}\left( t\right) $ have been obtained
 using the theory of quantum
integrals of motion (see Refs.\ \cite{DM89} and references therein).

\subsection{Bose-enhancement and normal density\label{SecNorDen}}

The effect of Bose enhancement is mathematically expressed as
exponentially growing solutions of the equations of motion for the
atomic field. Due to equivalence of the parametric approximation and
Hartree-Fock-Bogoliubov approach, Eqs.\ (\ref{HFB}) have in this case
exponentially growing solutions as well. With the neglect of
deactivating collisions and of depletion of the molecular field, and
for $\epsilon _{p}=$const, the equations for the anomalous and normal
 densities
attain the form of a system of inhomogeneous linear differential
equations,
\begin{eqnarray}
i \dot{m}_{s}\left( p,t\right) =2\epsilon _{p}m_{s}\left( p,t\right)
+2\tilde{g}^{*}\left\lbrack 1+2n_{s}\left( p,t\right) \right\rbrack
 \label{HFB0a}
\\
i \dot{n}_{s}\left( p,t\right) =2\tilde{g}^{*}m^{*}_{s}\left(
 p,t\right) -2\tilde{g}m_{s}\left( p,t\right)  . \label{HFB0n}
\end{eqnarray}
The corresponding homogeneous system has solutions $\sim \exp\left(
 \pm \kappa t\right) $. The
unstable modes with $\epsilon _{p}<2|\tilde{g}|$, or real $\kappa $
 [see Eq.\ (\ref{kappa})]
demonstrate exponential growth.

Some approximate approaches are based on the neglect of the normal
density $n_{s}\left( p,t\right) =0$. In this case Eq.\ (\ref{HFB0a})
 has only an oscillating
solution $\sim \exp\left( \pm 2i\epsilon _{p}t\right) $ and can not
 describe the effect of Bose enhancement. As
a result such approximations met little success in some situations
(see Ref.\
\cite{MSJ02}). The normal density have to be taken into account also
 for a
correct description of the squeezing of non-condensate atoms (see
 Sec.\
\ref{SecEntAt}). Nevertheless, the justified neglect of the normal
 density in
the first-order cummulant approach of Refs.\
\cite{KGB03,GKGB04,KGG04,GKGTJ04,GKB05} leads to significant
 simplification
of resulting equations and allows for the description of effects of
 spatial
inhomogeneity beyond the local density approximation. The normal
 density
could be taken into account in higher-order cummulant approaches.

\section{Loss of ultracold atoms\label{SecLoss}}

The effect of Feshbach resonance in ultracold atomic collisions has
been observed at first accompanied by a dramatic loss of trap
 population
when the magnetic field was tuned to the vicinity of the resonance.
 Such
loss has been observed both with atomic  BEC
 \cite{IASMSK98,SIAMSK99,C00}
and with ultracold thermal gases \cite{M02}.

These experiments can be generally divided to two categories, called
in the pioneer MIT works \cite{IASMSK98,SIAMSK99} as ''slow sweep''
 and
''fast sweep'' ones. In the fast sweep experiments the magnetic field
 has
been ramped through the resonance, while in the slow sweep ones the
 ramp
has been stopped short of crossing. The experiments of the second kind
use generally slower ramp speed than the fast sweep ones, or even a
static magnetic field, although the principal difference between the
 two
kinds of experiments is the presence or absence of resonance crossing.

The atomic loss is mainly attributed to two loss mechanisms. The
first one is a collision-induced deactivation process, such as
(\ref{AMCol}) and (\ref{MMCol}) \cite{TTHK99,YBJW99,YBJW00},
 undergone by
the temporarily formed molecular BEC. The second mechanism involves
 the
molecular dissociation induced by crossing to non-condensate atomic
states \cite{AV99,MTJ00,YBJW00}. The effects of the loss mechanisms
 are
non-additive (see Ref.\ \cite{YBJW00}). An additional loss can be
produced by secondary collisions of the relatively hot products of the
reactions (\ref{AMCol}) and (\ref{MMCol}) with condensate atoms (see
Refs.\ \cite{YBJW99,GS99,SMASRB01}).

\subsection{Deactivation loss mechanism}

This loss mechanism is included in the mean field approach
discussed in Sec.\ \ref{SecMeanField}. Certain properties of the loss
process can be described by analytical expressions, obtained from Eq.\
(\ref{deneq}), whenever the following ``{\it fast decay}'' conditions
 hold
(see Ref.\ \cite{YBJW00}): \begin{subequations} \label{FDA}
\begin{eqnarray}
&&\mu \dot{B}\ll \Gamma _{m}\left( D+\Gamma ^{2}_{m}/D\right) ,\quad
 \Gamma _{m}\gg \Gamma _{a}
\\
&&D^{2}+\Gamma ^{2}_{m}\gg 6|g|^{2}n_{0} . \label{FDAb}
\end{eqnarray}
\end{subequations}For definitions consult Eqs.\ (\ref{D}) and
(\ref{Gammaam}). These conditions mean that the relaxation of
 $n_{m}$, $v$,
and $u$ is much faster compared to that of $n_{0}$  and to the rate
 of change
of the energy, caused by the magnetic field with a ramp speed
 $\dot{B}$.
Therefore the values of the fast variables can be related to a given
value of the atomic condensate density $n_{0}$,  using a
 quasi-stationary
approximation, by
\begin{equation}
u\sim -{D\over \Gamma { } _{m}}v ,\qquad v\sim -{2|g
|^{2}n^{2}_{0}\Gamma { } _{m}\over D^{2}+\Gamma { } ^{2}_{m}} ,\qquad
 n_{m}\sim {|g|^{2}n{ } ^{2}_{0}\over D^{2}+\Gamma { } ^{2}_{m}} ,
 \label{quasist}
\end{equation}
and the condition (\ref{FDAb}) leads to $n_{m}\ll n_{0}$. As a
 result, a
single non-linear rate equation for the atomic density can be
extracted. When terms proportional to the atomic and molecular
densities in $D$ [see Eq.\ (\ref{D})] are neglected, the resulting
 rate
equation is
\begin{equation}
\dot{n}_{0}\left( {\bf r},t\right) =-{3|g|^{2}k_{a}n^{3}_{0}\left(
 {\bf r},t\right) \over \left\lbrack 2\epsilon _{a}\left( t\right)
+V\left( {\bf r}\right) \right\rbrack ^{2}+\left\lbrack
 k_{a}n_{0}\left( {\bf r},t\right) /2\right\rbrack { } ^{2}} .
 \label{RE}
\end{equation}
(The neglected terms in $D$ effectively add an extra shift to the
resonance, but its contribution is hardly noticed in the present
problem.)

Equation (\ref{RE}) has a form analogous to the Breit-Wigner
expression for resonant scattering in the limit of zero-momentum
collisions (see Refs.\ \cite{MTJ00,YB03b}).  In the Breit-Wigner sense
one can interpret $k_{a}n_{0}$  as the width of the decay channel,
 while the
width of the input channel is proportional to $|g|^{2}$. This
 observation
establishes a link between the macroscopic approach used here and
microscopic approaches that treat the loss rate as a collision
process. However, the right-hand side of Eq.\ (\ref{RE}) is four times
smaller than the usual Breit-Wigner expression. A factor of ${3\over
 2}$ is
associated with the loss of a third condensate atom in the reaction
(\ref{AMCol}), while an additional factor of ${1\over 6}$ is
 associated with the
effects of quantum statistics (see Refs.\ \cite{KSS85,B97}).

\subsubsection{Approaching the resonance\label{SecApprRes}}

Very close to resonance the behavior of Eq.\ (\ref{RE})
effectively attains a 1-body form linear in $n$. But as long as we
 stay
out of this narrow region, by obeying the ``{\it off-resonance}''
 condition
,
\begin{equation}
k_{a}n\left( {\bf r},t\right) \ll |V\left( {\bf r}\right) -\mu B\left
( t\right) | , \label{NRA}
\end{equation}
\noindent we can write Eq.\ (\ref{RE}) (to a very good approximation)
 in the
3-body form
\begin{equation}
\dot{n}_{0}=-3K_{3}\left( {\bf r},t\right) n^{3}_{0},\qquad
 K_{3}={2\pi |a_{a}|k_{a}\Delta \over m\mu \left\lbrack B\left(
 t\right) -B_{0}-V\left( {\bf r}\right) /\mu \right\rbrack { } ^{2}}
 . \label{K3}
\end{equation}
The three-body rate coefficient $K_{3}$  is defined here according to
conventional chemical notation. It is three times less then the one in
Refs.\ \cite{YBJW99,YBJW00,SIAMSK99}. The dependence of Eq.\
(\ref{K3}) on the scattering length $a_{a}$  follows from Eq.\
(\ref{Delta}).

The above results are obtained neglecting the atomic transport
described by the kinetic energy terms in Eq.\ (\ref{GPE}). This
approximation is valid while the characteristic times are small
compared to the trap periods. A different situation takes place in the
slow-sweep experiments \cite{IASMSK98,SIAMSK99}, where the
characteristic sweep times are large compared to the trap periods, and
the atomic transport has time to restore the Thomas-Fermi
distribution. In this case the averaging of Eq.\ (\ref{K3}) over the
distribution  (\ref{ThomasFermi}) leads to the following rate equation
\begin{equation}
\dot{\bar{n}}=-3{50\over 21}K_{3}\left( {\bf r},t\right) \bar{n}^{3}
\end{equation}
for the mean BEC density $\bar{n}\left( t\right) ={2\over
 5}n_{\text{peak}}\left( t\right) $.

\begin{figure}
\epsfxsize=0.7\textwidth  \epsfbox{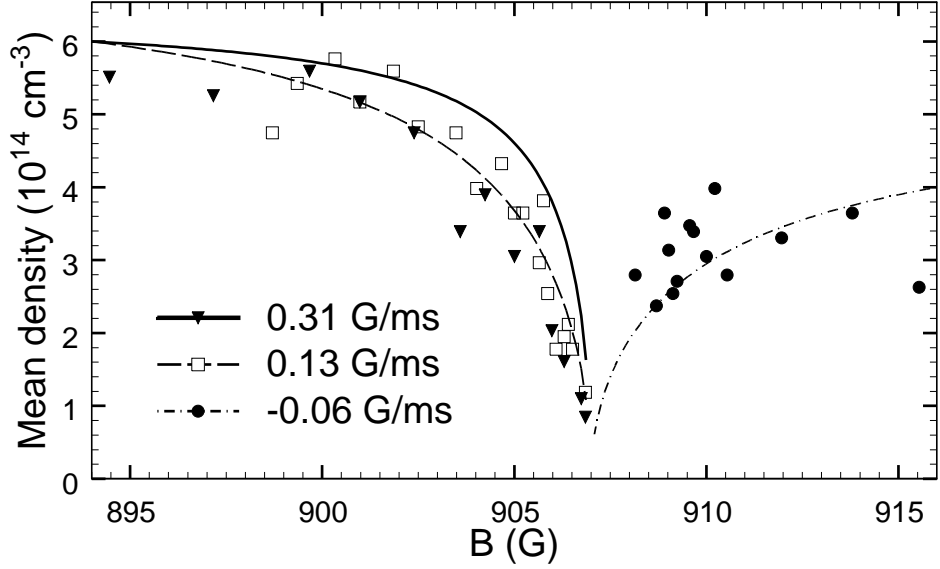}

\caption{The surviving mean density (see Eq.\
(\protect\ref{nt}) vs. the stopping value of the magnetic
field. The resonance was approached from below with two ramp
speeds, 0.13 G/s and 0.31 G/s, or from above, with the ramp
speed -0.06 G/s. These are compared with the experimental
results \protect\cite{SIAMSK99} (squares, triangles and
circles). \label{FigKet907n}}

\end{figure}

When the magnetic field ramp is assumed to vary linearly in time,
starting at $t_{0}$  and ending at $t$, and Eq. (\ref{NRA}) applies
throughout the ramp motion (i.e., by avoiding passage through the
resonance as in the slow sweep experiments), this rate equation can be
solved analytically as,
\begin{equation}
\bar{n}\left( t\right) =\bar{n}\left( t_{0}\right) \left\lbrack 1
+{200\over 7}\pi |a_{a}|k_{a}\bar{n}^{2}\left( t_{0}\right) \Delta
 {t-t{ } _{0}\over m\mu \dot{B}^{2}t_{0}t }\right\rbrack ^{-1/2},
 \label{nt}
\end{equation}
\noindent where $\dot{B}$ is the magnetic-field ramp speed and the
 extrapolated time of
reaching exact resonance is chosen to be $t=0$, so that both $t$ and
 $t_{0}$
have the same sign. A similar expression for the homogeneous case has
been presented in Refs.\ \cite{YBJW99,YBJW00}. Figure \ref{FigKet907n}
demonstrates a reasonably good agreement between the model and
experimental data of Ref.\ \cite{SIAMSK99} for the 907 G resonance in
$^{23}$Na. The model uses the parameter values $\Delta =0.98$ G,
 $a_{a}=3.4$ nm, and
$\mu =3.65 \mu _{B}$, calculated in Ref.\ \cite{MTJ00}, and
 $k_{a}=5.5\times 10^{-11}$cm$^{3}/$s,
measured in Ref.\ \cite{MAXCK04}, and does not contain adjustable
parameters.

\subsubsection{Passing through the resonance} \label{PasRes}

The off-resonance approximation of Eqs.\ (\ref{NRA}) and
(\ref{K3}) does not hold very close to resonance, and is therefore
inapplicable to the description of the fast-sweep experiment, in which
the Zeeman shift is swept rapidly {\it through} the resonance, causing
dramatic losses. Nevertheless, the fast decay approximation
(\ref{FDA}) may still be valid. A simple analytical expression can
then be derived for the condensate loss on passage though the
resonance if, in addition, the magnetic field variation lasts long
enough to reach the ``{\it asymptotic}'' condition
\begin{equation}
\mu \delta B\gg k_{a}n_{0} , \label{ascond}
\end{equation}
where $\delta B$ is the total change in $B$ accumulated over the
 sweep.
This condition allows the extension of the ramp starting and stopping
times to $\mp \infty $.

One can now evaluate the variation of $n_{0}\left( {\bf r},t\right) $
 in the infinite
time interval $\left( -\infty ,\infty \right) $. Let us rewrite Eq.\
(\ref{RE}) in the form
\begin{equation}
{dn_{0}\left( {\bf r},t\right) \over n^{2}_{0}\left( {\bf r},t\right)
 }=-{3k_{a}n_{0}\left( {\bf r},t\right) |g|{ } ^{2}\over \left( \mu
 \dot{B}t\right) ^{2}+\left( k_{a}n_{0}\left( {\bf r},t\right)
/2\right) { } ^{2}}dt ,
\end{equation}
where $V\left( {\bf r}\right) $ is removed by our choice of the
 origin on the time
scale. We then integrate the left-hand side with respect to $n_{0}$
from $n_{0}\left( {\bf r},-\infty \right) $ to $n_{0}\left( {\bf
 r},\infty \right) $ and the right-hand side with respect to $t$
from $-\infty $ to $\infty $, considering $n_{0}$  as a well-defined
 function of $t$. The
latter integral may be evaluated by using the residue theorem,
closing the integration contour by an arc of infinite radius in the
upper half-plane. The integral along this arc vanishes according to
the asymptotic behavior of $n_{0}$  (see Ref.\ \cite{YBJW00}).

The final result does not depend on $k_{a}$  and has the form (valid
for all positions ${\bf r}$)
\begin{equation}
n_{0}\left( {\bf r},\infty \right) ={n_{0}\left( {\bf r},-\infty
 \right) \over 1+s n_{0}\left( {\bf r},-\infty \right) },\qquad
 s={6\pi |g|{ } ^{2}\over \mu |\dot{B}|}={12\pi ^{2}|a_{a}|\over
 m}{\Delta \over |\dot{B}|} . \label{n}
\end{equation}
The product $s n_{0}$  in Eq.\ (\ref{n})  would be proportional to the
Landau-Zener exponent for the transition between the condensate and
the resonant molecular states whose energies cross due to the time
variation of the magnetic field, if one could keep the coupling
strength $g\varphi _{0}$  constant. However, for the non-linear curve
 crossing
problem represented by Eq.\ (\ref{GPE}), the Landau-Zener formula is
replaced by Eq.\ (\ref{n}), which predicts a lower crossing
probability, since the coupling strength $g\varphi _{0}$  decreases,
 along with
the decrease of the condensate density during the process. The
transition probability in the Landau-Zener model with relaxation,
obtained in Ref.\ \cite{DOS76}, is given by the same Landau-Zener
formula, as in the absence of the relaxation, and is independent of
the relaxation rate as well as Eq.\ (\ref{n}). However, unlike the
Landau-Zener formula, Eq.\ (\ref{n}) is applicable only at strong
relaxation, when the fast decay approximation is valid.

The asymptotic result (\ref{n}) describes the decay of the
condensate {\it density.} Assuming a homogeneous initial density
 within the
trap, Eq.\ (\ref{n}) applies also to the loss of the total {\it
 population}
$N_{0}\left( t\right) =\int n_{0}\left( {\bf r},t\right) d^{3}r$. An
 asymptotic expression for the total population
can also be found when the homogeneous distribution is replaced by the
Thomas-Fermi one [see Eq.\ (\ref{ThomasFermi})]. In this case, given
$n_{\text{peak}}$  is the maximum initial density in the center of
 the trap, one
obtains
\begin{equation}
{N_{0}\left( \infty \right) \over N_{0}\left( -\infty \right)
 }={15\over 2sn{ } _{\text{peak}}}\left( {1\over 3}+{1\over sn{ }
 _{\text{peak}}}-{1\over 2sn{ } _{\text{peak}}}\sqrt{1+{1\over sn{ }
 _{\text{peak}}}}\ln{\sqrt{1+{1\over sn{ } _{\text{peak}}}}+1\over
 \sqrt{1+{1\over sn{ } _{\text{peak}}}}-1}\right)  . \label{gd}
\end{equation}
\begin{figure}
\epsfxsize=0.5\textwidth  \epsfbox{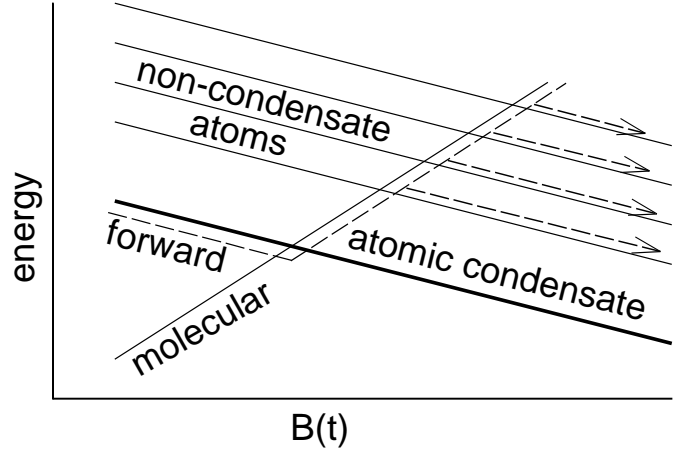}

\caption{Schematic illustration of transitions between
atomic and molecular condensates and non-condensate atoms on
a forward sweep. \label{FigForCross}}

\end{figure}

\subsection{Dissociation loss\label{SecCrossLoss}}

The consideration above did not include the decay of the resonant
molecular state by transferring atoms to excited (discrete) trap
 states or
to higher-lying non-trapped (continuum) states. In a forward sweep,
 when
the molecular state crosses the atomic ones upwards (see Fig.\
\ref{FigForCross}), this process can be described by curve-crossing
 theory
as a sequence of two-state crossings.

\begin{figure}
\epsfxsize=0.7\textwidth  \epsfbox{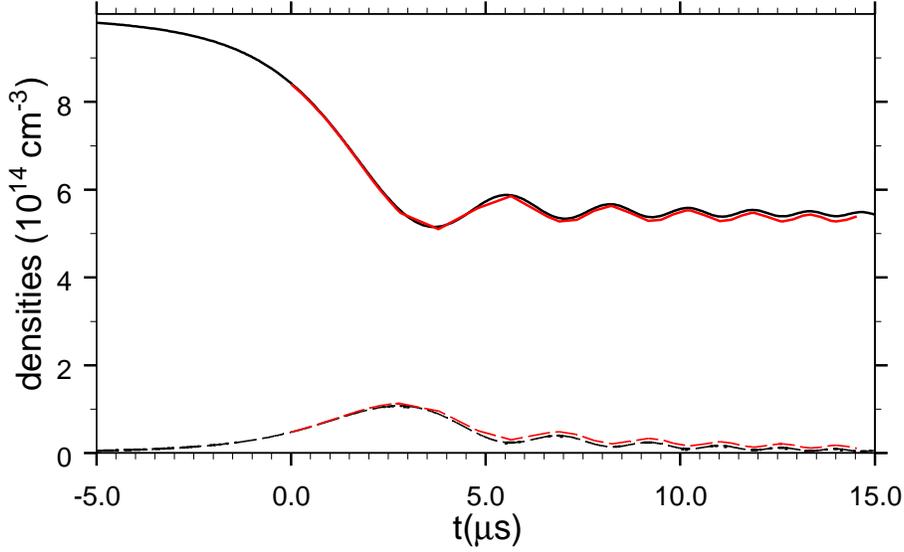}

\caption{Time dependence of the densities of the atomic
(solid lines)  and molecular (dashed lines) condensates
calculated for the Na resonance at 853 G using the
parametric approximation (black lines) and the mean-field
approach taking into account the dissociation width $\Gamma
 _{\text{dis}}$
(red lines). \label{FigCompMF}}

\end{figure}

Consider the system in a normalization box of volume ${\cal V}$. The
number of states of a pair of atoms with momenta ${\bf p}$ and $-{\bf
 p}$ per unit of
the energy $\epsilon =p^{2}/m$ of their relative motion is
\begin{equation}
\nu _{\epsilon }={{\cal V}\over 4\pi { } ^{2}}m^{3/2}\epsilon ^{1/2} .
\end{equation}
Due to the time variation of the magnetic field, the
molecular state crosses at the time $t$ the atomic pair state with
the energy $\epsilon =-2\epsilon _{a}\left( t\right) =\mu \left(
 B\left( t\right) -B_{0}\right) $, altogether crossing $\nu
 _{\epsilon }\mu \dot{B}$ such
states per unit time. If the depletion of the molecular field
during each crossing can be neglected, the crossings can be
described by the many-body curve-crossing  model of Sec.\
\ref{SecExSol}. The number of atoms formed by each crossing is
determined by Eq.\ (\ref{spd0}), with
\begin{equation}
\lambda ={8\pi |a_{a}|\Delta \over m|\dot{B}|}n_{m} . \label{lambdaNm}
\end{equation}
Therefore, the loss rate of the molecular population is
\begin{equation}
\dot{N}_{m}=-{1\over 2}\left( 1-e^{-2\pi \lambda }\right) e^{2\pi
 \lambda }\nu _{\epsilon }\mu \dot{B} . \label{Nmdot}
\end{equation}
If the molecular density is small enough, such that $\lambda \ll 1$,
 one
obtains a loss rate of the molecular density
\begin{equation}
\dot{n}_{m}\approx -2\Gamma _{\text{dis}}n_{m} ,\qquad \Gamma
 _{\text{dis}}=|a_{a}\mu |\Delta \sqrt{m\mu \left( B\left( t\right)
 -B_{0}\right) } , \label{CrossLoss}
\end{equation}
which is proportional to the dissociation width $\Gamma
 _{\text{dis}}$. The same
result can be obtained by using of Landau-Zener theory (see Ref.\
\cite{YBJW00}). Therefore one can account for the decay of the
resonant molecular state into excited trap states by adding a term
$-i\Gamma _{\text{dis}}\varphi _{m}$  to the right hand side of Eq.\
(\ref{GPEm}). This approach
is in good agreement with the parametric approximation only whenever
(a) the variation of the molecular field during each crossing is
negligible; and (b) $\lambda \ll 1$ and the quantum effect of Bose
 enhancement,
described by the positive-exponential factor in Eq.\ (\ref{Nmdot}), is
negligible (see Secs.\ \ \ref{SecExSol} and \ref{SecNorDen}). These
conditions are obeyed with the parameters used in the calculations of
Ref.\ \cite{YBJW00} for the Na resonances , the results of which are
confirmed by using the parametric approximation. For example, Fig.\
\ref{FigCompMF} compares the two approximate methods for the Na
resonance at 853 G with the strength $\Delta \approx 9.5$ mG (see
 Ref.\ \cite{MTJ00},
other parameters having the same values as for the 907 G resonance
presented at the end of Sec. \ref{SecApprRes}).

\begin{figure}
\epsfxsize=0.7\textwidth  \epsfbox{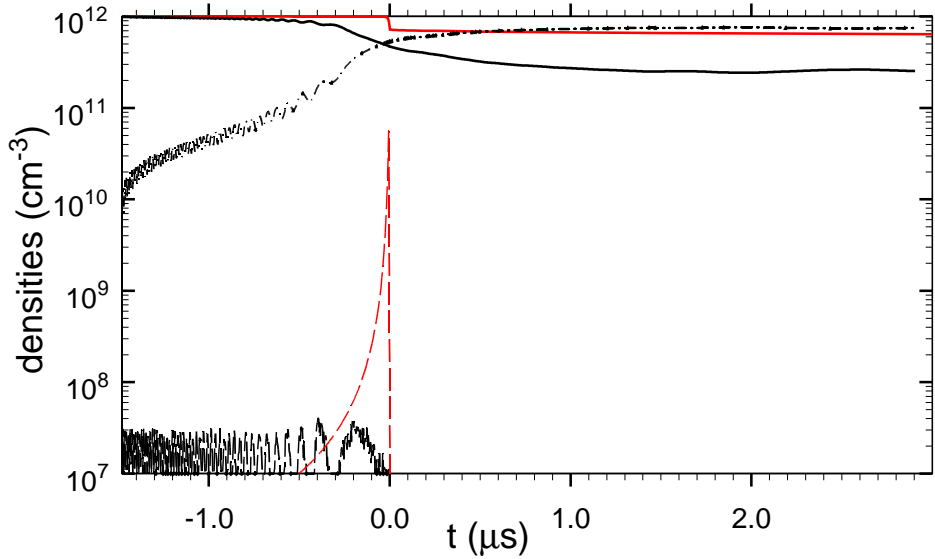}

\caption{Time dependence of the densities of the atomic (solid
lines)  and molecular (dashed lines) condensates calculated for the
$^{85}$Rb resonance using the parametric approximation (black lines)
 and
the mean-field approach taking into account the dissociation width
$\Gamma _{\text{dis}}$  (red lines). The dot-dashed line plots the
 density of non-
condensate atoms calculated using the parametric approximation. The
ramp speed is $\dot{B}=5$ G/ms.\label{FigRb85td}}

\end{figure}

\begin{figure}
\epsfxsize=0.8\textwidth  \epsfbox{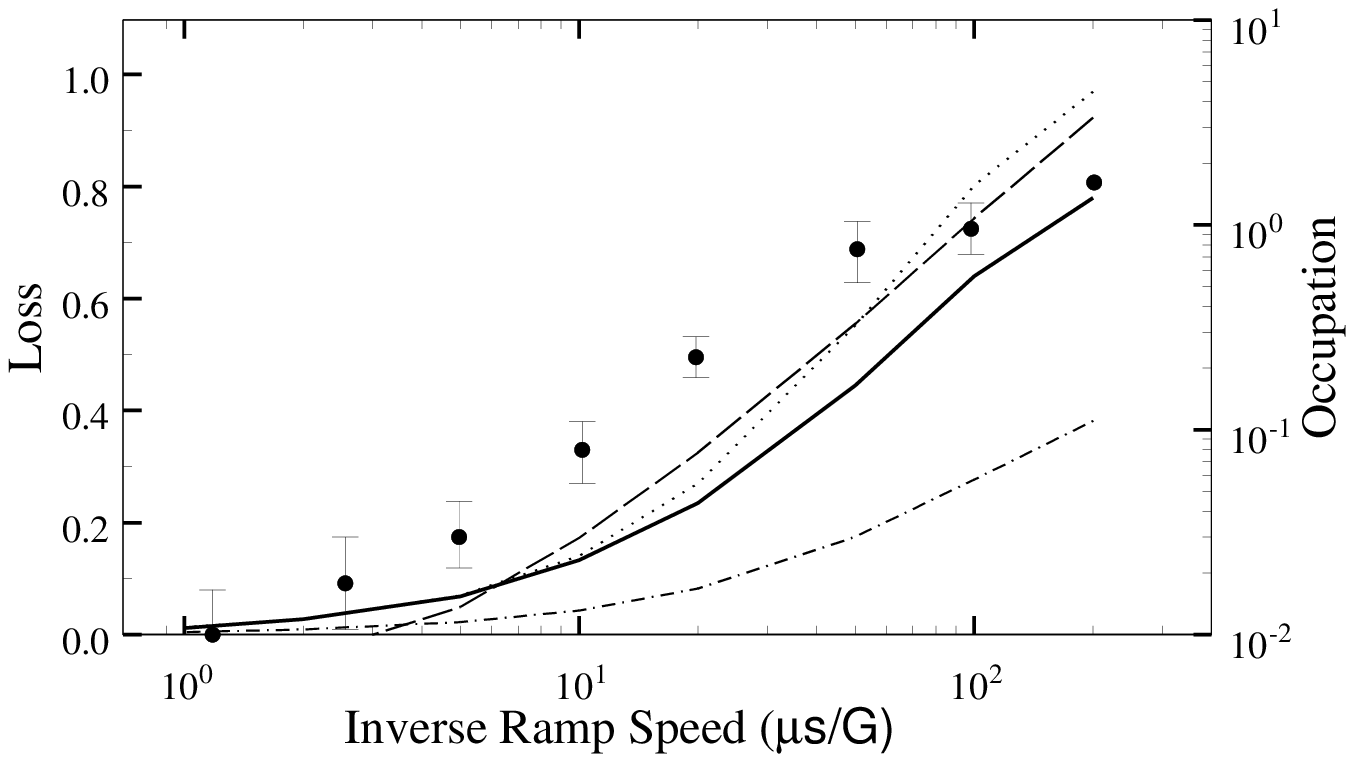}

\caption{Loss of atomic condensate (solid line) and
maximal non-condensate state occupation (dashed line) as a
function of inverse ramp speed calculated using the
parametric approximation in comparison with the experimental
data by Cornish {\it et al.} \protect\cite{C00} (circles).
The dot-dashed and dotted lines demonstrate, respectively,
the results of mean-field calculations taking into account
the dissociation width $\Gamma _{\text{dis}}$  and without this
 account for
$k_{a}=2\times 10^{-9}$  cm$^{3}/$s. \label{FigRb85loss}}

\end{figure}

However, under other conditions the two methods produce different
results. Figure \ref{FigRb85td} compares them for a case of $^{85}$Rb
resonance (see Ref.\ \cite{YB03j}). The values of $\Delta  =11$ G,
 $B_{0}=154.9$ G,
and $a_{a}=-450$ (in atomic units) are taken from Ref.\ \cite{D02},
 and the
value of $\mu  =-2.23 \mu _{\text{B}}$  is taken from Ref.\
 \cite{KH02}. The condensate
loss calculated with the parametric approximation is more then two
times higher then the one calculated with the mean field approach. The
main loss mechanism is the Bose-enhanced crossing to the
 non-condensate
atomic states. The Bose-enhanced dissociation drastically reduces the
molecular condensate density, compared to the mean field results,
leading to the suppression of deactivation losses (the total density
 of
condensate and non-condensate atoms remaining almost unchanged and the
results are insensitive to deactivation rates for $k_{a}, k_{m}
<10^{-8}$cm$^{3}/$s). As
a result, the parametric approximation demonstrates a much better
agreement with the experimental data than the mean field results (see
Fig.\ \ref{FigRb85loss}). The drastic increase of loss at slower
 sweeps
is actually related to the effect of Bose-enhancement, as demonstrated
by the plot for the maximal value of the non-condensate state
occupation $n_{s}\left( p,t\right) $  in Fig.\ \ref{FigRb85loss}. The
 disagreement
between the parametric approximation and the experimental data at
intermediate ramp speeds can be related to fluctuations of the
molecular field and higher-order correlations of the atomic field
neglected in both theories.

It is interesting that the inclusion of the dissociation loss
mechanism reduces the total loss (cf. the dotted and dot-dashed lines
 in
Fig..\ \ref{FigRb85loss}). This paradoxical result has the following
explanation. The coupling to non-condensate atom continuum shifts the
molecular state energy (above the continuum threshold the shift is
imaginary and proportional to the width). The energy-dependent shift
effectvely accelerates the sweep, reducing the probability of
 crossing from
the atomic to molecular BEC. For weaker resonances, when the crossing
probability remains almost unchanged, an additional effect becomes
 most
prominent (see Ref.\ \cite{YBJW00}). The atom-molecule deactivation
(\ref{AMCol}) leads to the loss of three condensate atoms per each
 resonant
molecule formed, while the dissociation leads to the loss of only two
condensate atoms. Therefore, the inclusion of dissociation losses
transfers flow from the deactivation (\ref{AMCol}) to the
 dissociation,
thus reducing the number of condensate atoms lost per each resonant
molecule formed. By the same reason in a case of small dissociation
 loss,
the increase of molecule-molecule deactivation rate coefficient
 reduces the
total loss (see calculations for Na in Ref.\ \cite{YBJW00}).

\begin{figure}
\epsfxsize=0.5\textwidth  \epsfbox{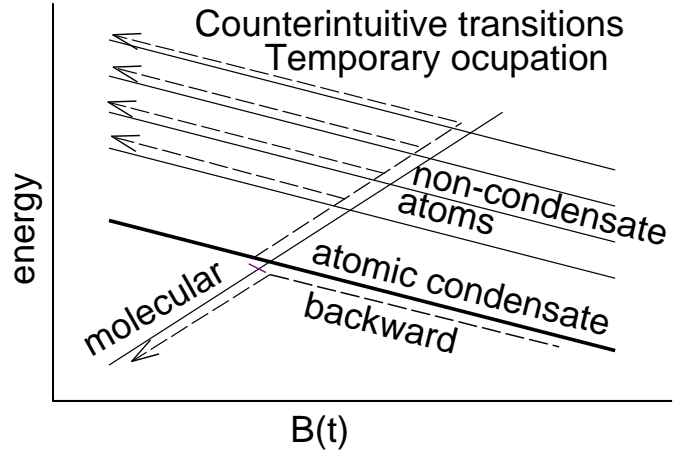}

\caption{Schematic illustration of transitions between
atomic and molecular condensates and non-condensate atoms on
a backward sweep. \label{FigCrossMol}}

\end{figure}

\section{Formation of molecules\label{SecForMol}}

In the experiments
\cite{HKMWCNG03,MKHCNG04,DVMR03,DVMR04,XMACMK03,MAXCK04} diatomic
molecules have been formed by sweeping the Zeeman shift through
resonance in a backward direction, so that the molecular state crossed
the atomic ones downwards (see Fig.\ \ref{FigCrossMol}). This led to
 the
transfer of population from the lowest atomic (condensate) state to
 the
molecular state, as had been proposed in Ref.\ \cite{MTJ00}. In a
backward sweep, unlike the forward one (see Fig.\ \ref{FigForCross}),
the transitions to non-condensate atomic states are forbidden in
semiclassical Landau-Zener-type theories (see Refs.\
\cite{ChildMCT,NU84,DO88,Akulin05}) as the second crossing, from the
molecular to a non-condensate atomic state, precedes in time the first
one, from the atomic to the molecular condensate state. Nevertheless,
quantum theory allows the non-condensate states to be populated
temporarily by so-called ``{\it counterintuitive transitions}'' (see
 Ref.\
\cite{counter_int}), which are most notable in strong resonances or at
low densities (see Ref.\ \cite{YB03}). The molecules are also lost due
to deactivating collisions with atoms or other molecules (\ref{AMCol})
and (\ref{MMCol}). Therefore, an adequate analysis of molecular
formation requires a quantum many-body theory that takes into account
relaxation, such as the parametric approximation.

\begin{figure}
\epsfxsize=0.7\textwidth  \epsfbox{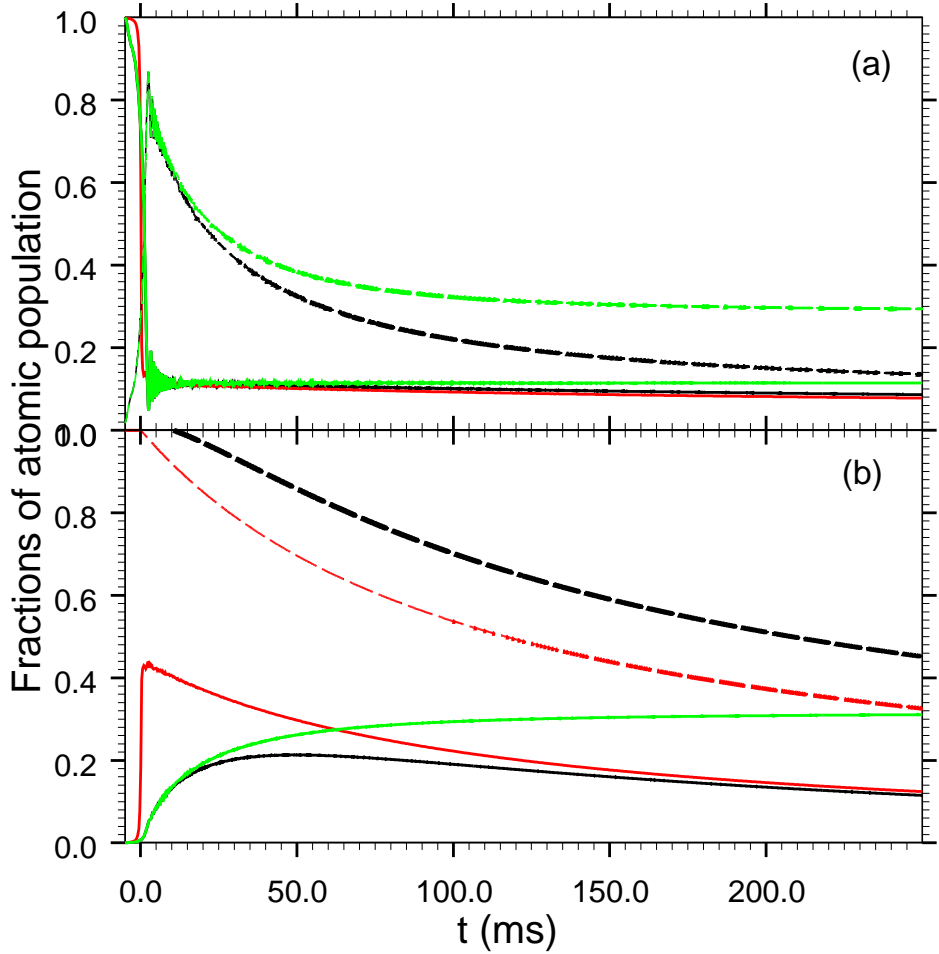}

\caption{Results of calculations for the 1007 G resonance in $^{87}$Rb
using the parametric approximation with deactivation (black lines) and
without it (green lines). The red lines plot the mean-field results.
Part (a) presents the fractions of the atomic populations surviving in
the condensate form (solid lines) and converted to non-condensate
atoms (dashed lines). The fractions of atomic population converted to
molecules are plotted by the solid lines in the part (b), while the
dashed lines plot the sum of all atomic and twice molecular
populations. All plots are calculated for the initial atomic density
$10^{11}$  cm$^{-3}$  and the ramp speed of 1 G/s in a backward sweep.
\label{FigRb1007m}}

\end{figure}

Figure \ref{FigRb1007m} compares various approximations for the
case of the 1007 G resonance in  $^{87}$Rb. The parameter values
 $\Delta =0.2$ G
and $a_{a}=99$ atomic units have been measured in Ref.\
 \cite{VDEMR03},
$\mu =2.8 \mu _{B}$  has been calculated in Ref.\ \cite{M02}, and
 $k_{a}=7\times 10^{-11}$
cm$^{3}/$s has been estimated in Ref.\ \cite{YB03b}. The
 molecule-molecule
deactivation rate coefficient is estimated here as
 $k_{m}=10^{-10}$cm$^{3}/$s.

\begin{figure}
\epsfxsize=0.7\textwidth  \epsfbox{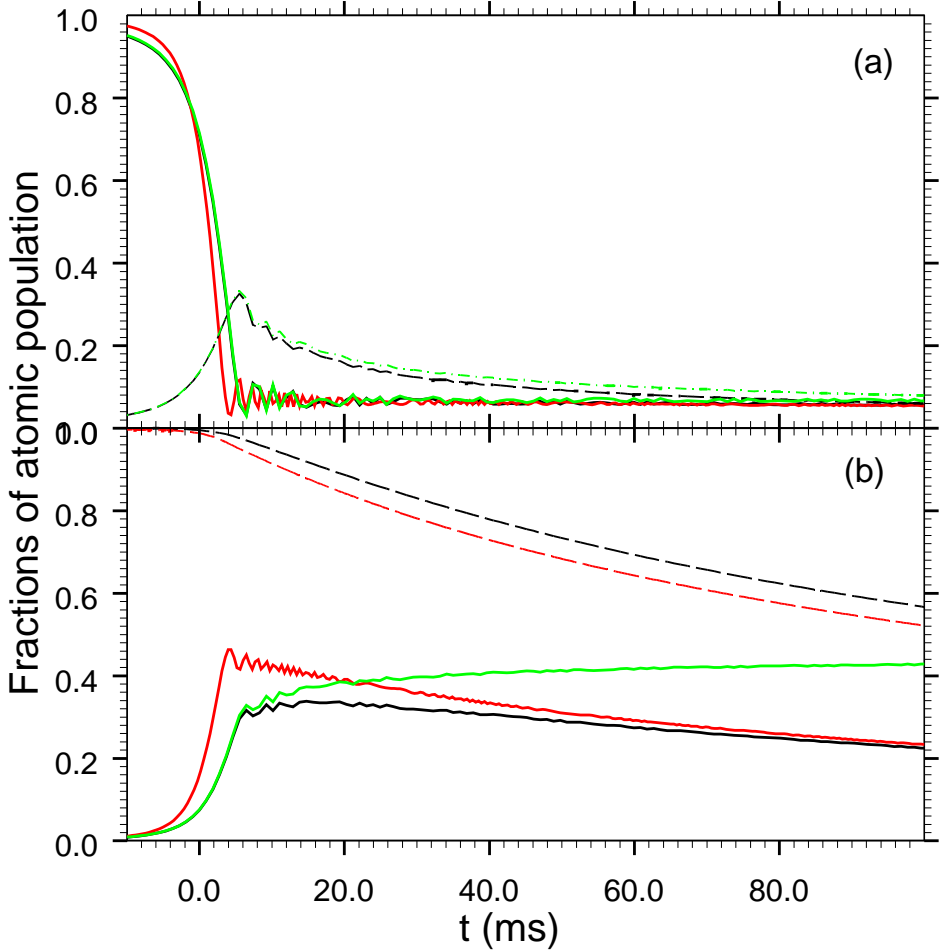}

\caption{Same as in Fig.\ \protect\ref{FigRb1007m}, but for the
685 G resonance in $^{87}$Rb and the ramp speed of 0.05 G/s.
\label{FigRb685m}}

\end{figure}

The results of calculations using the parametric approximation
demonstrate that the temporary non-condensate atom population persists
during a time comparable to the deactivation time. This population
substantially reduces the maximal molecular density and shifts the
maximum time compared to the mean-field results. This difference is
less pronounced in the case of the weaker 685 G resonance (see Fig.\
\ref{FigRb685m}), with $\Delta =17$ mG, $\mu =1.4 \mu _{\text{B}}$
(see Ref.\ \cite{M02}, other
parameters are the same as for the 1007 G resonance).

Figures \ref{FigRb1007m} and \ref{FigRb685m} contain
also results of parametric calculations neglecting the
deactivation losses ($k_{a}=k_{m}=0$). In this case the molecular
density increases monotonically in time due to the
association of non-condensate atoms, while the atomic
density remains almost unchanged.

The conversion efficiency $2\max\left( n_{m}\left( t\right) \right)
/n_{0}\left( t_{0}\right) $ is determined by a
concurrence of three processes: the association of the atomic
 condensate
and the two loss processes --- the dissociation of the molecular BEC
 onto
non-condensate atoms and the deactivation by inelastic collisions. A
qualitative behavior of the conversion efficiency can be analyzed
 using the
rescaled equation Eq.\ (\ref{Phimrd}). The first term in the
 right-hand
side is related to association, the second one --- to dissociation,
 and the
rest --- to deactivation. The problem depends on the parameter
 $\sigma $ [see Eq.\
(\ref{sigma})], the scaled deactivation rate coefficients
 $\tilde{k}_{a,m}$, and a
scaled ramp speed, which can be expressed as $\left( sn\right) ^{-1}$
  in terms of the
parameter $s$ defined in Eq.\ (\ref{n}). Given the rescaled ramp
 speed, an
increase of the resonance strength $\Delta $ increase the parameter
 $\sigma $ [see Eq.\
(\ref{sigma})] and, therefore, the dissociation term in Eq.\
(\ref{Phimrd}), but decrease the deactivation rates proportional to
 $\tilde{k}_{a,m}$
[see Eq.\ (\ref{kamresc})]. A decrease of the atomic density has the
 same
effect.

The part of associated atomic condensate increases with decrease
of the ramp speed, saturating at values of $sn\sim 1$ [see Eq.\
(\ref{n})
and the following discussion]. The molecular loss is proportional to
the time and, therefore, it is inverse proportional to the ramp speed.
This leads to the existence of an optimal ramp speed $\dot{B}\propto
 n\Delta $.

\begin{figure}
\epsfxsize=0.7\textwidth  \epsfbox{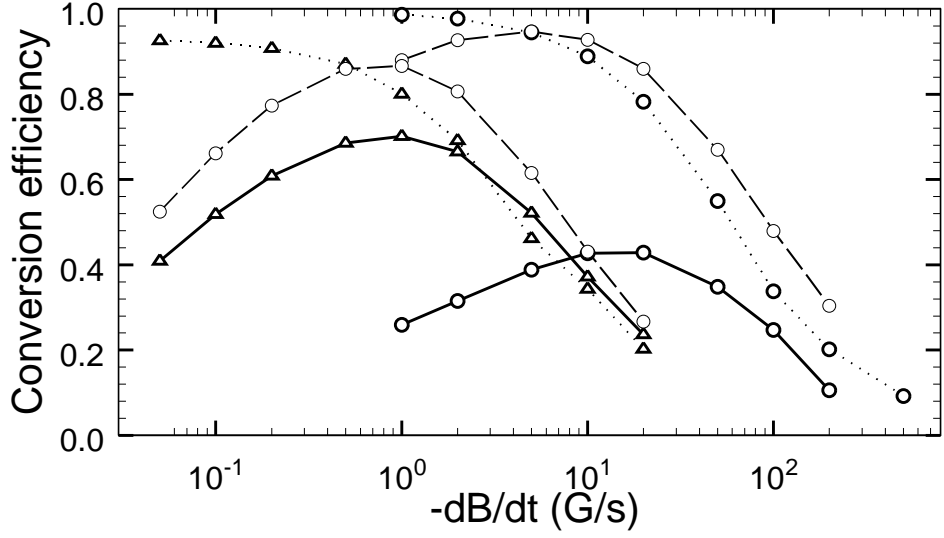}

\caption{Conversion efficiency as a function of the ramp speed
calculated for the 1007 G (circles) and 685 G (triangles) resonances
in  $^{87}$Rb with the initial atomic density of $10^{12}$  cm$^{-3}$
   using the
parametric approximation with deactivation (solid lines) and without
it (dotted lines). The dashed lines plot the mean-field
results.\label{FigPmb}}

\end{figure}

\begin{figure}
\epsfxsize=0.7\textwidth  \epsfbox{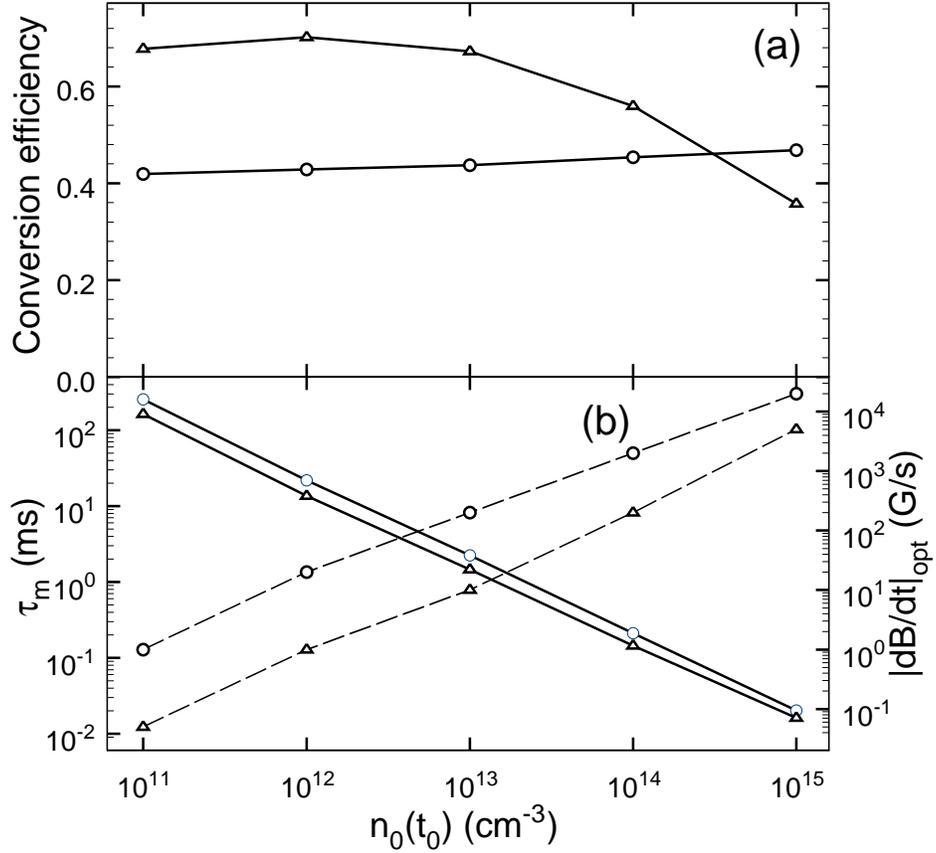}

\caption{(a) conversion efficiency and (b) the lifetime of the
molecular condensate $\tau _{m}$  (solid lines) at the optimal ramp
 speed
$\left( dB/dt\right) _{\text{opt}}$  (dashed lines), all  plotted as
 functions of the initial
atomic density $n_{0}\left( t_{0}\right) $, calculated with the
 parametric approximation,
for the resonances at 1007 G (circles) and 685 G (triangles) in
$^{87}$Rb.\label{FigPmn0}}

\end{figure}

This qualitative analysis is confirmed by results of numerical
calculations (see Figs.\ \ref{FigPmb} and \ref{FigPmn0}). The
maximum in the ramp-speed dependence is not reproduced by
calculations neglecting the deactivation. The mean-field
calculations predict the maximum, but overestimate the conversion
efficiency and provide a slower optimal ramp speed, especially for
the stronger resonance (see Fig.\ \ref{FigPmb}). The conversion
efficiency at the optimal ramp speed decreases at high densities due
to deactivation and at low densities due to non-condensate atom
population (see Fig.\ \ref{FigPmn0}), reaching a maximum at the
density $10^{12}$  cm$^{-3}$  for the weaker 685 G resonance. In the
 case of
the stronger 1007 G, resonance the non-condensate atom population is
the dominant loss process in the full range of densities considered
here, and the conversion efficiency remains almost unchanged,
slightly increasing with the density.

\begin{figure}
\epsfxsize=0.7\textwidth  \epsfbox{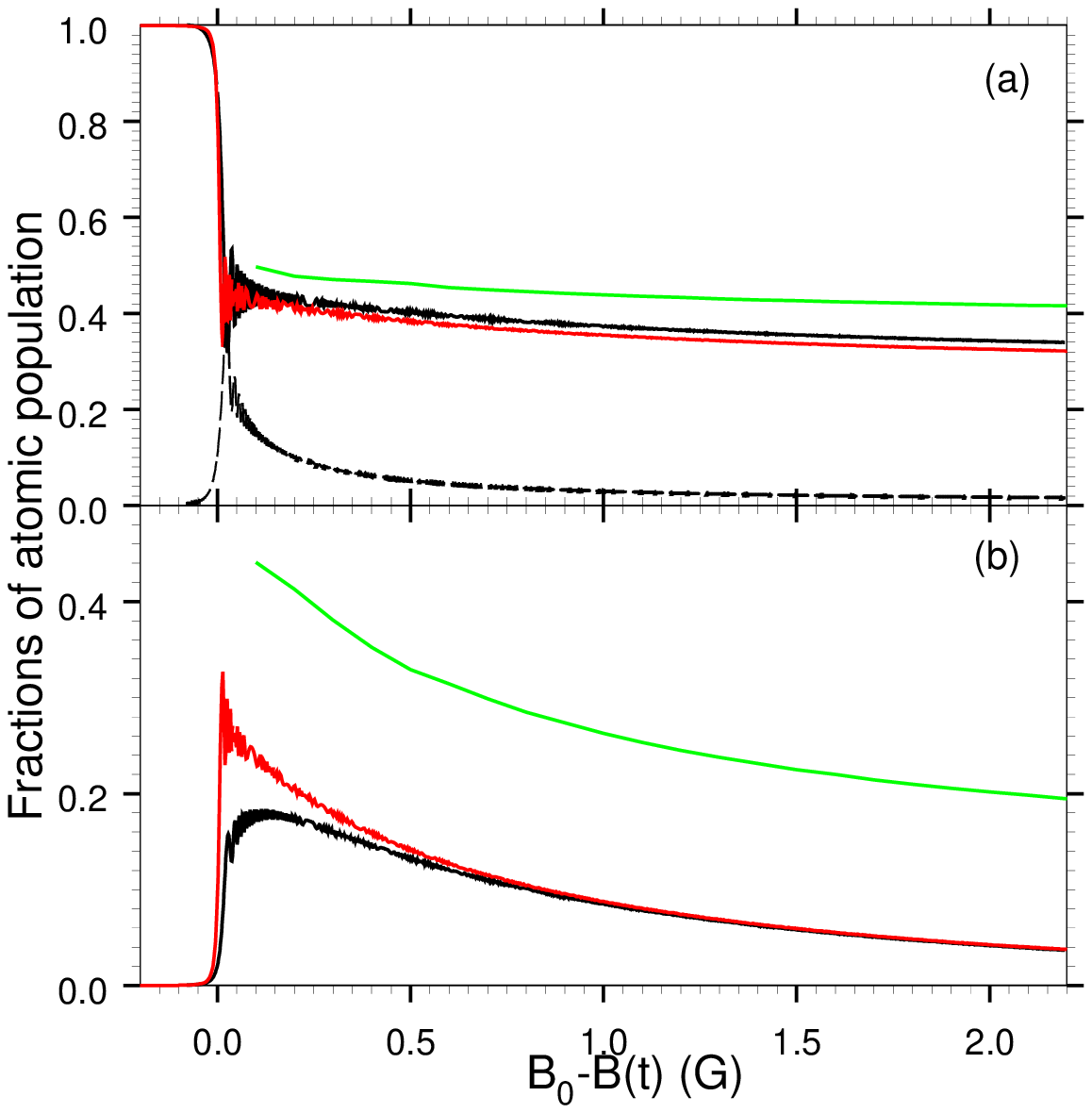}

\caption{Fraction of atomic population surviving in the atomic
condensate (a) and converted to molecules (b), calculated using the
parametric approximation (black lines) and the mean field approach
(red lines), for the 1007 G resonance in a homogeneous $^{87}$Rb BEC
 with
an initial atomic density $2\times 10^{13}$cm$^{-3}$ and a magnetic
ramp speed of 1 G/ms. The fraction of non-condensate atoms calculated
with the parametric approximation is represented by the dashed line in
part (b). The green lines represent the fraction of surviving and
converted population calculated using the mean field approach while
taking account of spatial inhomogeneity and expansion.\label{FigMFPA}}

\end{figure}

The low density required for a more effective conversion can be
achieved by the use of expanding condensates, as in experiments
\cite{HKMWCNG03,MKHCNG04,DVMR03,DVMR04}. In the $^{87}$Rb experiments
\cite{DVMR03,DVMR04}, the magnetic field ramp was started after a
preliminary expansion interval $t_{p}\ge 2$ ms, following the
 switching off of the
trap. In this case the analysis can be satisfactorily carried out by
 using
the mean-field approach (see Ref.\ \cite{YB04} and Sec.\
 \ref{SecExpan}).
This is demonstrated by Fig.\ \ref{FigMFPA}, comparing the atomic and
molecular densities calculated with the parametric approximation to
 those
calculated with the mean field approach, under the appropriate
 conditions.
The initial atomic density used corresponds to the mean density
 reached at
the expansion time of 2 ms. As one can see, already when the magnetic
 field
is just 0.5 G below the resonance (in a sweep totaling 2.2 G), the
 results
of the two kinds of calculations coincide. When faster magnetic
 sweeps or
higher densities are considered, the results of the two approaches
 converge
even faster.

The applicability of the mean-field approach at the experimental
conditions allows us to analyze the case of an expanding condensate
 using a
numerical solution of Eqs.\ (\ref{GPexpa}) and (\ref{GPexpm}) (see
 Fig.\
\ref{FigMFPA}). The expansion reduces the atomic density, leading to a
slower loss of atoms and molecules. The results are in agreement with
 the
experimental data of Ref.\ \cite{DVMR03} reporting that $\sim 7\%$ of
 the
atoms are converted to molecules and $\sim 30\%$ remain in the atomic
condensate for ramp speeds of less than 2 G/ms.

\section{Formation of entangled atoms\label{SecEntAt}}

The dissociation of the molecular BEC in a forward sweep (see
Fig.\ \ref{FigForCross}) has been considered in Sec.\ \ref{SecLoss} as
one of the loss mechanisms. However, the dissociation is not just an
undesirable process, since the molecules dissociate to entangled pairs
of atoms with opposite momenta (see Ref.\ \cite{YB03}), as in the case
of an unstable atomic BEC (see Ref.\ \cite{Y02}). Indeed, substitution
of the molecular mean field (\ref{Psiphim}) into the atom-molecule
coupling (\ref{dhm}) leads to an interaction proportional to a product
of atomic field operators with opposite momenta. Interactions of
similar form are known in the theory of  parametric down conversion in
quantum optics (see Refs.\ \cite{SZ97,BR97}). Likewise, the atoms are
formed in two-mode squeezed states, which are similar to the state of
electromagnetic radiation formed by parametric down conversion.

In order to clarify the nature of this entanglement, let us write
out a pseudo-Hamiltonian
\begin{equation}
\hat{H}_{a}=\sum\limits^{}_{p_{z}>0}\hat{H}\left( {\bf p},t\right)  ,
\end{equation}
that leads to equations of motion for the atomic field
(\ref{Psia}), expressed as a sum of contributions of different
momentum modes in a normalization box,
\begin{eqnarray}
\hat{H}_{a}\left( {\bf p},t\right) =\left\lbrack {p{ } ^{2}\over 2m}
+ \epsilon _{a}\left( t\right) -i{k{ } _{a}\over 2}|\varphi _{m}\left
( t\right) |^{2}\right\rbrack \left\lbrack \hat{\Psi }^{\dag
 }_{a}\left( {\bf p},t\right) \hat{\Psi }_{a}\left( {\bf p},t\right)
+\hat{\Psi }^{\dag }_{a}\left( -{\bf p},t\right) \hat{\Psi }_{a}\left
( -{\bf p},t\right) \right\rbrack  \nonumber
\\
+\biggl\{2\tilde{d}_{h}\left( 2p\right) \varphi ^{*}_{m}\left(
 t\right) \hat{\Psi }_{a}\left( {\bf p},t\right) \hat{\Psi }_{a}\left
( -{\bf p},t\right) +i \hat{F}\left( {\bf p},t\right) \left\lbrack
 \hat{\Psi }^{\dag }_{a}\left( {\bf p},t\right)  + \hat{\Psi }^{\dag
 }_{a}\left( -{\bf p},t\right) \right\rbrack  + \mathrm{h.c.}\biggr\}
 . \nonumber
\end{eqnarray}
Although the pseudo-Hamiltonian is not hermitian, it allows
writing the time evolution operator in the form
\begin{equation}
\hat{\Omega }\left( t\right) =\prod^{}_{p_{z}>0}\hat{\Omega }\left(
 {\bf p},t\right)  ,
\end{equation}
where (using the time-ordering operator $T$)
\begin{equation}
\hat{\Omega }\left( {\bf p},t\right) =T \exp\left(
 -i\int\limits^{t}_{t{ } _{0}}\hat{H}_{a}\left( {\bf p},t^\prime
 \right) d t^\prime \right) .
\end{equation}
The representation of the operator $\hat{\Omega }\left( t\right) $ as
 a product of
single-mode operators $\hat{\Omega }\left( {\bf p},t\right) $ follows
 from the commutativity of the
$\hat{H}_{a}\left( {\bf p},t\right) $ with different values of ${\bf
 p}$.

Imagine a measurement, represented by a projection operator
 $\hat{P}\left( {\bf p}\right) $,
which selects atoms with the momentum ${\bf p}$ moving in the positive
$z$-direction, and does not affects atoms moving in the negative
$z$-direction. This measurement reduces the state vector $\hat{\Omega
 }\left( t\right) |$in$\rangle $ to
$\hat{P}\left( {\bf p}\right) \hat{\Omega }\left( t\right)
|$in$\rangle =\hat{\Omega }\left( {\bf p},t\right) |$in$\rangle $.
 The distribution of atoms moving in the
negative $z$-direction (the average number of atoms with the momentum
${\bf p}^\prime $) after this measurement will be determined by
\begin{equation}
\langle \text{in}|\hat{\Omega }^{\dag }\left( t\right) \hat{\Psi
 }^{\dag }_{a}\left( {\bf p}^\prime ,t\right) \hat{\Psi }_{a}\left(
 {\bf p}^\prime ,t\right) \hat{\Omega }\left( {\bf p},t\right)
|\text{in}\rangle \propto  \delta _{-{\bf p}{\bf p}^\prime } ,
\end{equation}
representing an entanglement of atoms with opposite momenta. This
analysis is similar to the one used in the entanglement of the signal
and the idle in the process of degenerate two-photon down-conversion
in quantum optics (see Refs.\ \cite{SZ97,BR97}). The state of atoms is
also perfectly number-squeezed, i. e. it has zero variance in the
relative number of atoms with opposite momenta, as in the case
considered in Ref.\ \cite{RGB02}.

As demonstrated in  \cite{YBJ02} the non-condensate atoms
produced by molecular dissociation are formed in squeezed states,
which now turn out to be two-mode squeezed states, as in  \cite{Y02}.
As in quantum optics, the squeezing is related to the quadrature
operators
\begin{equation}
\hat{X}\left( {\bf p},t\right) ={1\over 2}\Bigg\{\left\lbrack
 \hat{\Psi }_{a}\left( {\bf p},t\right) \pm \hat{\Psi }_{a}\left(
 -{\bf p},t\right) \right\rbrack e^{i\theta }+\left\lbrack \hat{\Psi
 }^{\dag }_{a}\left( {\bf p},t\right) \pm \hat{\Psi }^{\dag
 }_{a}\left( -{\bf p},t\right) \right\rbrack e^{-i\theta }\Biggr\} .
 \label{Quadrature}
\end{equation}
The uncertainties of the quadratures can be written out as
\begin{equation}
\langle \text{in}|\hat{X}\left( {\bf p}_{1},t\right) \hat{X}\left(
 {\bf p}_{2},t\right) |\text{in}\rangle ={1\over 2}\delta \left( {\bf
 p}_{1}-{\bf p}_{2}\right) \left\{1+2n_{s}\left( p,t\right) \pm
 2\text{Re}\left\lbrack m_{s}\left( p,t\right) e^{2i\theta
 }\right\rbrack \right\}
\end{equation}
where the momentum spectra $n_{s}\left( p,t\right) $ and $m_{s}\left(
 p,t\right) $ are defined by
Eqs.\ (\ref{ns}) and (\ref{Anden}), respectively.

The uncertainties attain maximal and minimal values at two
orthogonal values of the phase angle $\theta $. The amount of
 squeezing can
be quantified by the energy-dependent parameter
\begin{equation}
r\left( \epsilon ,t\right) ={1\over 4}\ln{\langle \text{in}
|\hat{X}\left( {\bf p}_{1},t\right) \hat{X}\left( {\bf p}_{2},t\right
) |\text{in}\rangle { } _{\max}\over \langle \text{in}|\hat{X}\left(
 {\bf p}_{1},t\right) \hat{X}\left( {\bf p}_{2},t\right)
|\text{in}\rangle { } _{\min}}={1\over 4}\ln{1+2n_{s}\left( p,t\right
) +2|m_{s}\left( p,t\right) |\over |1+2n_{s}\left( p,t\right) -2
|m_{s}\left( p,t\right) ||}. \label{re}
\end{equation}
A mean squeezing parameter, weighed by the spectral density of
Eq.\ (\ref{nsE}),
\begin{equation}
\bar{r}\left( t\right) =\int dE \tilde{n}_{s}\left( E,t\right) r\left
( E,t\right) /n_{s}\left( t\right) . \label{avsq}
\end{equation}
is used to describe the time variation of the squeezing.

\begin{figure}
\noindent
\epsfxsize=\textwidth  \epsfbox{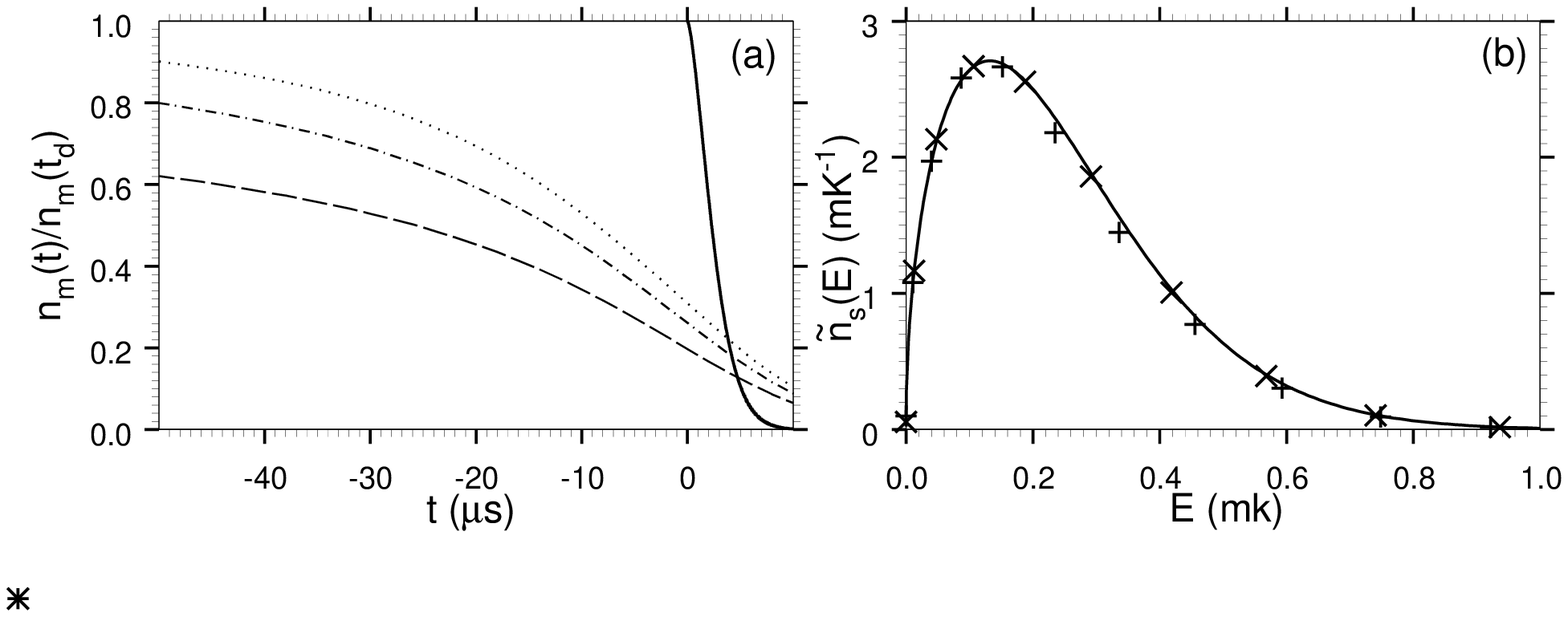}

\caption{(a) Fraction of surviving molecular population
calculated by using Eq.\ (\protect\ref{Nmsurv}) (solid line) and by
the parametric approximation for the initial molecular density
$n_{m}\left( t_{d}\right) =10^{10}$  cm$^{-3}$(dotted line),
 $10^{11}$  cm$^{-3}$(dot-dashed line), and
$10^{12}$  cm$^{-3}$(dashed line). (b) Energy distribution calculated
 by using
Eq.\ (\protect\ref{Edist}) (solid line) and the parametric
approximation for the initial molecular density $10^{11}$  cm$^{-3}$
($\times $) and
$10^{12}$  cm$^{-3}$(+).\label{FigDiss}}

\end{figure}

In the case of a rather weak resonance a description of the
dissociation of a pure molecular BEC can be attempted by a simple
analytical model based on the approach of Sec.\ \ref{SecCrossLoss}
(see also Refs.\ \cite{MAXCK04,GKGTJ04,YB04}). The molecular density
is given by the solution of Eq.\ (\ref{CrossLoss}),
\begin{equation}
n_{m}\left( t\right) =n_{m}\left( t_{d}\right) \exp\left\lbrack -2
|a_{a}\mu |\Delta \int\limits^{t}_{t{ } _{d}}dt^\prime \sqrt{m\mu
 \left( B\left( t'{}\right) -B_{0}\right) }\right\rbrack ,
 \label{Nmsurv}
\end{equation}
where $t_{d}$  is the time of start of the dissociating ramp. A
molecule dissociates into two entangled atoms, each with an energy
 $E$,
by a crossing occurring at $\mu \left( B\left( t\right)  -B_{0}\right
)  =2E$. The resulting energy
distribution of the formed atoms can be written out as
\begin{equation}
\tilde{n}_{s}\left( E\right) ={1\over 2}\sqrt{E}E^{-3
/2}_{\text{peak}}\exp\left\lbrack -{1\over 3}\left( E
/E_{\text{peak}}\right) ^{3/2}\right\rbrack , \label{Edist}
\end{equation}
where
\begin{equation}
E_{\text{peak}}={1\over 2m}\left( {m|\dot{B}|\over 4|a_{a}|\Delta
 }\right) ^{2/3} \label{Epeak}
\end{equation}
is the peak energy. Although the model does not produce the
correct molecular density dynamics (see Fig.\ \ref{FigDiss}a), it
gives a surprisely good agreement with the numerical results for the
energy distribution (see Fig.\ \ref{FigDiss}b). The analytical model
demonstrates also a very good agreement to experimental data, as
reported in Refs.\ \cite{MAXCK04,DVMR04}. It should be noted that at
high energy Eqs.\ (\ref{ns}), (\ref{EnergySpect}), and (\ref{psicsas})
lead to $\tilde{n}_{s}\left( E\right) \propto E^{-3/2}$. Therefore,
 the distribution has a divergent mean
energy. The peak energy (\ref{Epeak}) is well defined for each
distribution. It is proportional to the mean energy for the
distribution (\ref{Edist}) used in  Refs.\ \cite{MAXCK04,DVMR04}.

In the case of molecular condensate dissociation the squeezing
parameter does not exceed the value of $r=0.35$ even for the initial
molecular density of $10^{12}$  cm$^{-3}$. The nature of such low
 squeezing can
be understood by using the exactly soluble many-body curve-crossing
model of Sec.\ \ref{SecExSol}. Using expressions for the uncertainties
of the quadrature operators (\ref{Quadrature}) derived in Ref.\
\cite{YBJ02}, one can represent the squeezing parameter (\ref{re}) as
\begin{equation}
r={1\over 2}\ln{\left( e^{2\pi \lambda }-1\right) ^{1/2}+e{ } ^{\pi
 \lambda }\over |\left( e^{2\pi \lambda }-1\right) ^{1/2}-e^{\pi
 \lambda }|} , \label{rlambda}
\end{equation}
where the Landau-Zener parameter $\lambda $ is defined by Eq.\
(\ref{lambdaNm}). In the case of $\lambda \ll 1$, when Eqs.\
(\ref{Nmsurv}) and
(\ref{Edist}) are applicable [see discussion following to Eq.\
(\ref{CrossLoss})], the squeezing parameter is $r\approx \sqrt{2\pi
 \lambda }\ll 1$. In the
opposite case of $\lambda \gg 1$, Eq.\ (\ref{rlambda}) relates the
 squeezing
parameter to the occupation of the non-condensate mode $n_{s}\left(
 p,t\right) \approx e^{2\pi \lambda }$
[see Eq.\ (\ref{spd0}] as $r={1\over 2}\ln\left( 4n_{s}\right) $, in
 agreement with an exact
relation $n_{s}=\sinh^{2}r$, derivable from the definition of
 squeezed states
(see Refs.\ \cite{SZ97,BR97}). This relation between high squeezing
and high mode occupation demonstrates that a correct calculations of
the entangled atomic pair properties require taking into account the
normal density.

\begin{figure}
\noindent
\epsfxsize=\textwidth  \epsfbox{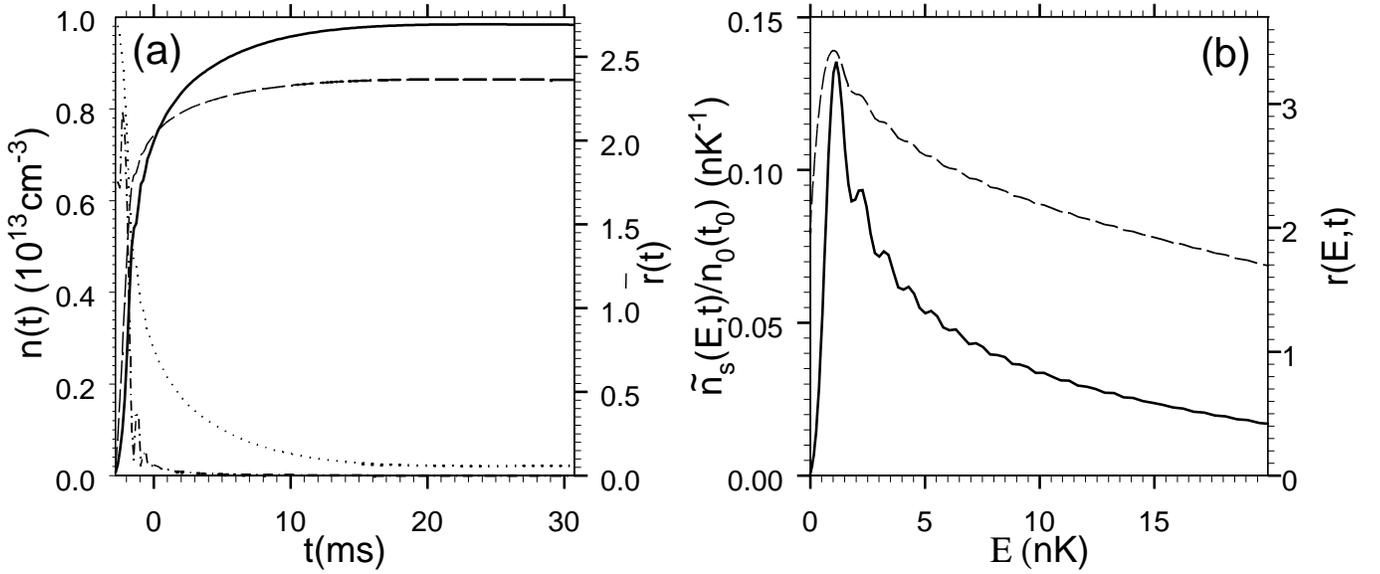}

\caption{(a) Time dependence of the densities of the atomic
condensate (dotted line), entangled atoms (solid line), and the
molecular condensate (dot-dashed line, $\times 10^{3}$) calculated
 for the 685 G
resonance in $^{87}$Rb with the initial atomic density $10^{13}$
 cm$^{-3}$   and a
ramp speed of 20 G/s in a forward sweep. The dashed line shows the
mean squeezing parameter $\bar{r}\left( t\right) $ [see Eq.\
(\protect\ref{avsq})]. (b)
Energy spectra of the entangled-atom density $\tilde{n}_{s}\left(
 E,t\right) $ (solid line) and
the squeezing parameter $r\left( E,t\right) $ (dashed line)
 calculated at $t\approx 23$ ms.
\label{FigRb685sq}}

\end{figure}

Higher squeezing can be obtained on dissociation of molecules formed
temporarily from the atomic BEC on a forward sweep (see Fig.\
\ref{FigForCross}). In this case (see Fig.\ \ref{FigRb685sq}) the
 molecular
density is very low (about $10^{10}$  cm$^{-3}$) and persists a
 shorter time (compared
to that in the backward sweep) due to fast dissociation, leading to a
negligibly small deactivation loss. Figure \ref{FigRb685sq}a
 demonstrates
almost full transformation of the atomic BEC into a gas of entangled
 atomic
pairs in two-mode squeezed states with the mean squeezing parameter
 $\bar{r}\approx 2.4$,
corresponding to a noise reduction of about 20 dB. The energy spectra
 of the
entangled-atom density and the squeezing parameter are presented in
 Fig.\
\ref{FigRb685sq}b. The density spectra are rather narrow, and the peak
energy increases with time. The squeezing parameter reaches the value
 of
$r\left( E,t\right) \approx 3.4$ (corresponding to noise reduction of
 about 30 dB) for the mode
with occupation $n_{s}\approx 240$ at the energy $E\approx 1$ nK.
 Comparable squeezing of $\max
r\left( E,t\right) \approx 3.6$ have been calculated in Ref.\
 \cite{YB03} for the 853 G resonance
in Na. These values substantially exceed $\max r\left( E,t\right)
 \approx 1.4$ calculated in Ref.\
\cite{YB03j} for the much stronger resonance in $^{85}$Rb.

\section*{Conclusion}

A correct theoretical treatment of Feshbach resonance in BEC has to
take into account both quantum fluctuation (formation of
 non-condensate
atoms) and damping due to deactivation of the resonant molecules in
inelastic atom-molecule and molecule-molecule collisions. The
 deactivating
collisions lead to condensate loss and the limitation of the
 atom-molecule
conversion efficiency at high densities and slow sweeps. The quantum
fluctuations of the atomic field describe Bose-enhanced condensate
 losses,
the limitation of atom-molecule conversion efficiency at low
 densities, and
the formation of entangled atomic pairs in two-mode squeezed states.
 The
effects of fluctuations and deactivation on the loss are
 non-additive. The
squeezing and Bose-enhancement of non-condensate atom formation can be
described by second-order correlations, but only if both normal and
anomalous densities are taken into account.

The parametric approximation takes into account both the
damping and fluctuations in a proper way. This method gives
analytical solutions for some model systems and can be used for
numerical calculations in more general cases. The results of
numerical calculations generally agree with the experimental data
and provide optimal conditions for atom-molecule conversion and
quantum properties of entangled atoms.

\section*{Acknowledgments}

The author is most grateful to Abraham Ben-Reuven for
invaluable and continuous collaboration in development and
applications of the theoretical methods presented in the chapter.

\end{document}